\documentclass[prd,aps,showpacs,tightenlines,preprint,nofootinbib,floatfix]{revtex4-1} 

\usepackage{epsfig}
\usepackage{graphicx}
\usepackage{multirow} 
\usepackage{color}
\usepackage[dvipsnames]{xcolor}
\usepackage{hyperref}
\usepackage{amsmath,amssymb}
\usepackage{rotating}
\usepackage[caption=false]{subfig}
\usepackage{slashed}
\usepackage{york}
\newcommand{\diracslash}[1]{#1\!\!\!\!/}
\begin{document}
\preprint{CERN-TH-2018-207}

\title{Dark Matter from Strong Dynamics: \\The Minimal Theory of Dark Baryons}

\newcommand\york{Department of Physics and Astronomy, York University, Toronto, Ontario, M3J 1P3, Canada}

\newcommand\cern{Theoretical Physics Department, CERN, CH-1211 Geneva 23, Switzerland}

\author{Anthony~Francis$^1$\email{anthony.francis@cern.ch}}
%\affiliation{\cern}
\author{Renwick~J.~Hudspith$^2$\email{renwick.james.hudspith@googlemail.com}}
%\affiliation{\york}
\author{Randy~Lewis$^2$\email{randy.lewis@yorku.ca}}
%\affiliation{\york}
\author{Sean~Tulin$^2$\email{stulin@yorku.ca}}
%\affiliation{\york}

\affiliation{$^1$\cern,$^2$\york}

\date{\today}

\begin{abstract}

As a simple model for dark matter, we propose a QCD-like theory based on $\SU(2)$ gauge theory with one flavor of dark quark.
The model is confining at low energy and we use lattice simulations to investigate the properties of the lowest-lying hadrons.
Compared to QCD, the theory has several peculiar differences: there are no Goldstone bosons or chiral symmetry restoration when the dark quark becomes massless; the usual global baryon number symmetry is enlarged to $\SU(2)_B$, resembling isospin; and baryons and mesons are unified together in $\SU(2)_B$ iso-multiplets.
We argue that the lightest baryon, a vector boson, is a stable dark matter candidate and is a composite realization of the hidden vector dark matter scenario.
The model naturally includes a lighter state, the analog of the $\eta^\prime$ in QCD, for dark matter to annihilate into to set the relic density via thermal freeze-out. 
Dark matter baryons may also be asymmetric, strongly self-interacting, or have their relic density set via $3 \to 2$ cannibalizing transitions.
We discuss some experimental implications of coupling dark baryons to the Higgs portal.

\end{abstract}


\maketitle

%% introduction
\section{Introduction}
\label{sec:intro}

The mass and stability of luminous matter in the Universe are largely a byproduct of QCD. Around 99\% of the mass of baryonic matter arises from the strong interaction and its stability is a consequence of an accidental $\text{U}(1)_B$ baryon number symmetry in the Standard Model (SM). 
Yet baryons represent only about one sixth of the total mass budget of matter in the Universe. 
The remainder is dark matter (DM). 
Although unknown in its properties, many different models and detection strategies have been proposed for DM, often motivated by other issues associated with the SM (e.g., the stability of the weak scale or the strong CP problem)~\cite{Bertone:2004pz,Feng:2010gw}.

%Astrophysical observations have established the important role of DM throughout the Universe's evolution.
%However, these observations point only to DM's gravitational influence and its particle dynamics remain unknown.
%Since DM has no viable explanation within the SM, uncovering its particle theory represents one of the most important avenues for discovering physics beyond the SM. 

On the other hand, dark and luminous matter may come from two separate particle physics sectors, orthogonal to any problems of the SM~\cite{Strassler:2006im}. DM may be the lightest stable state within a dark sector that has its own gauge group and matter representations. 
If these fields are singlets under the SM gauge group, the interactions between the two sectors only arise through higher dimensional operators and may be very feeble. 
However, interactions {\it within} the dark sector are generically much larger, especially for a non-abelian gauge theory that is strongly coupled.
Moreover, long-standing puzzles on the galactic-scale structure of DM provide an astrophysical motivation for strong self-interactions between DM particles~\cite{Spergel:1999mh} (see Ref.~\cite{Tulin:2017ara} for a review).
Composite dark sectors are a natural framework for self-interacting DM~\cite{Cline:2013zca,Boddy:2014yra}.

It is an appealing hypothesis that the mass and stability of DM arise through strong dynamics, similar to luminous baryons. 
Early realizations along these lines include technicolor baryons~\cite{Nussinov:1985xr,Chivukula:1989qb,Barr:1990ca} and mirror baryons~\cite{Chacko:2005pe,Foot:2014mia}. 
For strongly coupled gauge theories, lattice field theory is the main calculational tool in the nonperturbative regime. 
While mainly used for QCD~\cite{Wilson:1974sk}, recent studies of non-abelian dark sectors have turned to the lattice to investigate the basic properties of these theories, such as the spectrum of states and form factors for interactions with the SM~\cite{Lewis:2011zb,Hietanen:2013fya,Appelquist:2013ms,Detmold:2014qqa,Detmold:2014kba,Appelquist:2014jch,Appelquist:2015yfa,Appelquist:2015zfa,Francis:2016bzf}. 
We refer the reader to Ref.~\cite{Kribs:2016cew} for a recent survey of different models along these lines. 

In this work, we propose a minimal model realizing these ideas and compute its basic properties on the lattice.
The DM candidate in our model will be the lightest baryon in a strongly coupled Yang-Mills theory, which is stable due to an accidental symmetry. 
By minimal, we mean the fewest number of colors $N_c$ and flavors $N_f$, and the smallest nontrivial representation for matter fields. 
Hence, we consider $\SU(2)$ gauge theory with one Dirac fermion $q$ (dark quark) in the fundamental representation.
We do not consider the case of a single Weyl fermion due to Witten's anomaly~\cite{Witten:1982fp}. 
Hambye and Tytgat proposed a similar DM model based on $\SU(2)$ gauge theory with scalar quarks~\cite{Hambye:2009fg}.

In the space of gauge theories, $N_c\! = \! 2$ theories have long been useful as a simplified version of QCD~\cite{Marinari:1981nu,Kogut:1983ia,Nakamura:1984uz,vonSmekal:2012vx}. 
However, an important distinction is the fact that the fundamental representation of $\SU(2)$ is pseudo-real, unlike in $\SU(N_c)$ with $N_c > 2$. 
As a consequence, two-color theories have an enlarged global symmetry that reflects transformations between quarks and antiquarks. 
At the hadron level, there is a unification of baryons, antibaryons, and mesons.

In our one-flavor theory, the quark $q$ and antiquark $\bar{q}$ fields form a doublet, written schematically as $\mathcal{Q} \sim \left( \begin{smallmatrix} q \\ \bar{q} \end{smallmatrix} \right)$. 
As we show in Sec.~\ref{sec:model}, the theory has an unbroken global $\SU(2)_B$ symmetry acting on $\mathcal{Q}$.\footnote{Henceforth, $\SU(2)_B$ denotes the global symmetry, while $\SU(2)$ without the subscript refers to the local gauge symmetry.}
This symmetry is a non-abelian generalization of a $\text{U}(1)_B$ baryon number symmetry for $q$; it is clear that $\text{U}(1)_B$ is the diagonal subgroup of $\SU(2)_B$. 
Since the whole setup is analogous to isospin, we refer to this symmetry as {\it baryonic isospin}.
We argue below that $\SU(2)_B$ is not violated by chiral symmetry breaking or a finite mass for $q$.
Hence, the hadronic spectrum of the theory will fall nicely into $\SU(2)_B$ iso-multiplets. 

We envision that the lightest baryon in our theory will be a suitable DM candidate. The lightest $qq$ state is part of a spin-1 iso-triplet
\begin{equation}
\label{eq:rhomultiplet}
\rho = \left( \begin{array}{c} \rho^+ \\ \rho^0 \\ \rho^- \end{array} \right) 
\sim \left( \begin{array}{c} qq \\ \tfrac{1}{\sqrt 2}(q \bar{q} + \bar{q} q ) \\ \bar{q}\bar{q} \end{array} \right) \, .
\end{equation}
Borrowing an analogy from QCD, this state is akin to the $\rho$ meson. 
However, the superscripts in Eq.~\eqref{eq:rhomultiplet} refer not to electric charge, but to $\text{U}(1)_B$ charges: baryon ($+$), antibaryon ($-$), and meson ($0$). 
All three components are stable DM candidates provided $\SU(2)_B$ remains unbroken. 

Another peculiar feature of our $N_f=1$ model is the absence of Goldstone bosons. 
Once the axial $\text{U}(1)_A$ anomaly is considered, no chiral symmetries are present in the ``chiral'' limit, where $q$ becomes massless (see discussion in Ref.~\cite{Creutz:2006ts}). 
The would-be Goldstone boson from the $\text{U}(1)_A$ symmetry, which we denote $\eta$ (analogous to the $\eta^\prime$ in QCD), acquires a mass through the anomaly. 
This stands in contrast to $\SU(2)$ gauge theory with $N_f = 2$, which has an enlarged pion sector compared to QCD and the lightest baryons are themselves Goldstone bosons~\cite{Lewis:2011zb}.

In the early Universe, strong interactions in the dark sector populate a thermal plasma of dark quarks and gluons, which later are confined into hadrons after a cosmological phase transition, similar to QCD. 
The DM relic density may be frozen-out before or after the transition, depending on the dark quark mass $m_q$ and the confinement scale $\LambdaMS$. 
In the latter case ($m_q \lesssim \LambdaMS$), an appealing feature of our model is that there is a built-in annihilation channel $\rho \rho \to \eta \eta$ for setting the relic density provided $m_\rho > m_\eta$ (with the $\eta$ subsequently decaying into SM particles).
It is one of our key lattice results that this inequality holds for {\it any} value of $m_q$, unlike QCD where $m_\rho < m_{\eta^\prime}$.
Annihilation is important for standard freeze-out~\cite{Scherrer:1985zt} or asymmetric freeze-out where the dark sector has a dark baryon asymmetry~\cite{Petraki:2013wwa,Zurek:2013wia}.
The precise details depend on the relative temperature of the dark sector~\cite{Berezhiani:1995am}, its coupling with the visible sector (see, e.g.,~\cite{Kuflik:2015isi}), and the possible role of cannibalizing transitions~\cite{Carlson:1992fn,Choi:2017zww}, such as $\rho\rho\rho \to \rho \rho$.
We defer an analysis of the cosmology of our model to future study.

The remainder of this work is organized as follows. 
In Sec.~\ref{sec:model}, we present our dark sector model, including the leading non-renormalizable operators with SM fields.
We discuss the $\SU(2)_B$ symmetry properties of the Lagrangian and other bilinear operators that will be relevant for the lattice computations.
We also discuss recent arguments that $\SU(2)$ gauge theory does not provide a suitably stable DM candidate~\cite{Appelquist:2015yfa} and argue that the $\rho$ meson in our model avoids these pitfalls.
Section \ref{sec:setup} describes the lattice ensembles that we use, how quark propagators are constructed, and provides a first look at the hadron spectrum.
We devote particular attention to defining the ``chiral'' $m_q=0$ limit in this Goldstone boson-less theory.
We also compute $\LambdaMS$ as a convenient scale to normalize dimensionless lattice quantities into physical units.
Section \ref{sec:spectrum} presents our main results: the calculation of the dark hadron mass spectrum and decay constants.
In Sec.~\ref{sec:sigmaterm}, we discuss couplings between the dark sector and the SM and implications for DM detection. In particular, we use the Feynman-Hellman theorem to provide a determination of the Higgs coupling to our dark matter candidate.
Conclusions are provided in Sec.~\ref{sec:conclusions}.
The appendices describe two complementary methods to determine the ``chiral'' point, provide an alternative and more precise approach to defining physical scales, and summarize our lattice ensembles.

%% our pheno physics model
\section{Dark Sector Model}
\label{sec:model}

\subsection{Renormalizable Lagrangian and bilinear operators}

The Lagrangian for $\SU(2)$ gauge theory with one Dirac fermion $q$, with mass $m$, is
\begin{equation}
{\cal L} = -\tfrac{1}{2} \textrm{Tr}(F_{\mu\nu} F^{\mu\nu})+\bar{q}(i \diracslash{D} - m) q + \mathcal{L}_{\rm higher \; dim} \, .
\label{eq:lagrangian}
\end{equation}
We assume that $q$ is in the fundamental representation of $\SU(2)$ and is a singlet under the SM gauge symmetries. The covariant derivative is $D_\mu = \partial_\mu + \tfrac{i}{2} g A_\mu^a \sigma^a$, where $g$ is the gauge coupling and $\sigma^a$ represents the Pauli matrices acting on $\SU(2)$ color indices. Although there are no renormalizable interactions between the dark sector and the SM, the two sectors may couple through higher dimensional operators, which we discuss below.

By analogy with QCD, Eq.~\eqref{eq:lagrangian} has a $\U(1)_L \times \U(1)_R$ chiral symmetry for $m = 0$. 
However, the two-color theory is different from QCD since the fundamental representation of $\SU(2)$ is pseudo-real. 
Our theory possesses an enlarged $\text{U}(2)$ global symmetry.\footnote{For completeness, we mention that for a general $\SU(2)$ gauge theory with $N_f$ flavors, the usual $\U(N_f)_L \times \U(N_f)_R$ chiral symmetry is enlarged to $\U(2N_f) = \U(1)_A \times \SU(2N_f)$. Chiral symmetry breaking reduces $\SU(2N_f) \to \text{Sp}(2N_f)$ (i.e., the compact symplectic group), yielding $(2 N_f + 1)(N_f - 1)$ Goldstone bosons. For $N_f=1$, no Goldstones appear since $\SU(2) = \text{Sp}(2)$.}
To see this, we can write the fermion part of Eq.~\eqref{eq:lagrangian} in the following form
\begin{equation} \label{eq:Lferm}
\mathcal{L}_{\rm fermion} = \bar{\mathcal Q} i \diracslash{D} \mathcal{Q} 
- \tfrac{m}{2} \left( \mathcal{Q}^T i \sigma^2 C E \mathcal{Q} + \bar{\mathcal{Q}} i \sigma^2 C E \bar{\mathcal{Q}}^T \right) 
\end{equation}
where $C$ is the charge conjugation matrix acting on Dirac spinors and
\begin{equation}
\mathcal{Q} = \left( \begin{array}{c} q_L \\ -i \sigma^2 C \bar{q}_R^T \end{array} \right) \; , \qquad 
E = \left( \begin{array}{cc} 0 & 1 \\ -1 & 0 \end{array} \right) \, . 
\end{equation}
The kinetic term in Eq.~\eqref{eq:Lferm} is manifestly invariant under $\U(2)$ transformations acting on $\mathcal{Q}$.
For the mass term, let us decompose the global symmetry as $\U(2) = \U(1)_A \times \SU(2)_B$, since rotating $\mathcal{Q}$ by an overall phase is equivalent to an axial $\U(1)_A$ transformation on $q$. As mentioned in the introduction, $\SU(2)_B$ is a baryonic isospin symmetry, with $\U(1)_B$ as a subgroup, that plays a similar role as isospin in QCD. While $\U(1)_A$ is broken for $m\ne 0$, $\SU(2)_B$ remains intact since $E$ is an invariant tensor. 

In lattice calculations, local operators constructed from $q,\bar{q}$ create and annihilate states in the hadronic spectrum with the same quantum numbers. 
In this work, we consider states with $J^P = 0^\pm$ and $1^\pm$. 
The relevant mesonic operators are
\begin{subequations}
\label{eq:operators}
\begin{align}
&{\rm scalar\; (0^+)} & \mathcal{O}_S &= \bar{q} q 
= \tfrac{1}{2} \left( \mathcal{Q}^T i \sigma^2 C E \mathcal{Q} + \bar{\mathcal{Q}} i \sigma^2 C E \bar{\mathcal{Q}}^T \right)
\\
&{\rm pseudoscalar \; (0^-)} & \mathcal{O}_P &= \bar{q}\gamma_5 q 
= - \tfrac{1}{2} \left( \mathcal{Q}^T i \sigma^2 C E \mathcal{Q} - \bar{\mathcal{Q}} i \sigma^2 C E \bar{\mathcal{Q}}^T \right) 
\\
&{\rm vector \; (1^-)} & \mathcal{O}_V^\mu &= \bar{q}\gamma^\mu q 
= \bar{\mathcal{Q}} \gamma^\mu \tau^3 \mathcal{Q} 
\\
&{\rm axial \; vector \; (1^+)} &\mathcal{O}_A^\mu &= \bar{q}\gamma^\mu \gamma^5 q 
= \bar{\mathcal{Q}} \gamma^\mu \mathcal{Q}  \, .
\end{align} \end{subequations}
On the right-hand side, we have expressed these operators in terms of $\mathcal{Q}$ to make clear the $\SU(2)_B$ isospin properties of these states. 
The scalar, pseudoscalar, and axial vector operators are iso-singlets. 
To write the vector operator, we introduce Pauli matrices $\tau^a$ acting on isospin indices. The vector operator is part of an iso-triplet 
\begin{equation} \label{eq:Ovec}
\mathcal{O}_V^{a\mu} = \bar{\mathcal{Q}} \gamma^\mu \tau^a\mathcal{Q}
\end{equation}
that includes both meson and diquark operators. From Eq.~\eqref{eq:Ovec}, we see that the lightest baryon in the theory has $J^P=1^-$ and forms a triplet under $\SU(2)_B$, described in Eq.~\eqref{eq:rhomultiplet}.
We also write the tensor bilinear as
\begin{equation}
\mathcal{O}_T^{\mu\nu} = 
\bar{q} \sigma^{\mu\nu} q
= \mathcal{Q}^T E \tau^3 C \sigma^{\mu\nu} (i \sigma^2) \mathcal{Q} 
- \bar{\mathcal{Q}} E \tau^3 C \sigma^{\mu\nu} (i \sigma^2) \bar{\mathcal{Q}}^T \, ,
\end{equation}
where $\sigma^{\mu\nu} = \tfrac{1}{2}[\gamma^\mu,\gamma^\nu]$. Since a $\tau^3$ is required, it transforms under $\SU(2)_B$ like $\mathcal{O}_V^\mu$.

By analogy with QCD, we expect the chiral condensate $\langle \bar{q} q \rangle$ to receive a nonzero value through spontaneous symmetry breaking. 
However, since $\bar q q$ is an iso-singlet operator, its vacuum expectation value does not violate baryonic isospin. 
On the other hand, the chiral condensate breaks the global $\U(1)_A$ symmetry, potentially leading to a pseudo-Goldstone boson $\eta$ that becomes massless for $m=0$. 
However, just as in QCD, $\U(1)_A$ is anomalous, which gives an additional contribution to the $\eta$ mass.

\subsection{Nonrenormalizable interactions and CP violation}

The dark sector and SM may be coupled through higher-dimensional operators. 
The leading operators, arising at dimension five, are $\mathcal{O}_{S,P} |H|^2$, where $H$ is the SM Higgs field. 
When the Higgs field gets its vev $\langle H \rangle = v/\sqrt{2}$, there is an additional contribution to the dark quark mass.
In general, this term need not be aligned with the Dirac mass $m$ and there may be a relative CP-violating phase between them.\footnote{Our model has another source of CP violation from the $\theta$ term. For simplicity, we have neglected this term in the present work.} 
If we start in a basis where only $\mathcal{O}_S |H|^2$ appears, we must allow $m$ to be complex:
\begin{equation}
\mathcal{L} \supset - m\,  \bar{q}_R q_L - m^* \, \bar{q}_L q_R - \frac{1}{M} \bar{q} q |H|^2 \,  .
\end{equation}
Here, $M$ is the mass scale parametrizing the coupling between the two sectors. 
Performing a chiral rotation to make the total quark mass real and positive, we have
\begin{equation}
\mathcal{L} \supset - m_q\,  \bar{q} q  - \frac{1}{M} \left( \cos\phi \, \mathcal{O}_S + \sin\phi \, \mathcal{O}_P \right) \left( v h + \tfrac{1}{2} h^2  \right) \,  , \label{eq:higherdim}
\end{equation}
where $m_q = |m + \tfrac{1}{2M} v^2|$ and $\phi = \arg(m + \tfrac{1}{2M} v^2 )$.
CP violation manifests as a coupling between both operators $\mathcal{O}_{S,P}$ and the Higgs boson $h$.
The pseudoscalar coupling is particularly important since it causes the $\eta$ meson to be cosmologically unstable.
Phenomenological consequences of Eq.~\eqref{eq:higherdim} are explored in Sec.~\ref{sec:sigmaterm}.

\subsection{Dark matter stability}

Let us now discuss the question of whether the lightest baryon in our theory, the $\rho$ meson, provides a suitable DM candidate. 
Ref.~\cite{Appelquist:2015yfa} argued that if DM is stabilized by an accidental symmetry, the symmetry must be preserved including operators of dimension five, not just at the renormalizable level.
Dimension-five operators, even if suppressed by the Planck scale, may induce DM to decay much more rapidly than the age of the Universe.
On the other hand, dimension-six operators lead to a cosmologically acceptable DM lifetime if the suppression scale $M$ is large enough (but below the Planck scale).
This is the same situation as the proton in the SM: since the leading operators contributing to proton decay arise at dimension six, protons are cosmologically stable for $M \gtrsim 10^{13}$ GeV.
According to Ref.~\cite{Appelquist:2015yfa}, this argument disfavors $\SU(2)$ dark sectors since the global $\U(1)_B$ symmetry may be violated by dimension-five operators of the form $\sim qq |H|^2$.

However, these arguments do not apply to our $N_f=1$ model.
The only dimension-five operators are $\mathcal{O}_{S,P} |H|^2$ and neither allow for $\rho$ decay since they do not violate $\SU(2)_B$.
The leading operators that violate $\SU(2)_B$ must involve $\mathcal{O}_V^{a\mu}$ and arise at dimension six or higher by Lorentz symmetry. 
Therefore, we conclude that the $\rho$ meson is a viable DM candidate in terms of its stability, while iso-singlet states, such as the $\eta$ meson, are not.

In fact, the scale of physics connecting the dark and visible sectors need not be extremely high ($M \gg {\rm TeV}$) to preserve $\rho$ stability. 
For example, if the two sectors are coupled through a singlet scalar field, $\SU(2)_B$ is still preserved since the scalar may only couple to $\mathcal{O}_{S,P}$. 
Alternatively, if a $Z^\prime$ gauge boson mediates the coupling, it may couple to $\mathcal{O}_V^{a\mu}$. 
This will break the $\SU(2)_B$ down to its $\U(1)_B$ subgroup and, while the $\rho^0$ will be destablized, the $\rho^\pm$ remains a stable DM candidate. 
Hence, DM stability is robust in the face of these simplest mediators between sectors.

%% ensembles and propagators
\section{Lattice setup}
\label{sec:setup}

\subsection{Lattice ensembles and propagators}

For our lattice study, we discretize the one-flavour $\SU(2)$ theory of Sec.~\ref{sec:model} to arrive at the familiar Wilson action,
\begin{eqnarray}
S_W &=& \frac{\beta}{2}\sum_{x,\mu,\nu}\left(1-\frac{1}{2}{\rm ReTr}U_\mu(x)U_\nu(x+\mu)U_\mu^\dagger(x+\nu)U_\nu^\dagger(x)\right) + (4+m_0)\sum_x\bar\psi(x)\psi(x) \nonumber \\
&& -\frac{1}{2}\sum_{x,\mu}\bigg(\bar\psi(x)(1-\gamma_\mu)U_\mu(x)\psi(x+\mu)+\bar\psi(x+\mu)(1+\gamma_\mu)U_\mu^\dagger(x)\psi(x)\bigg)
\end{eqnarray}
where $U_\mu(x)$ is the $\SU(2)$ gauge field and $\psi(x)$ is the 4-component Dirac spinor for the dark quark.
The sum over $x$ covers the entire lattice and in this work we primarily use $V=L^3\times T=12^3\times32$, but for certain topics we will use $12^3\times48$ lattices also. We choose to use this arguably small volume for our exploratory study compared to those of typical lattice QCD simulations as we want to cover a large range of bare input masses $m_0$ on reasonable resources. For the light quark spectrum, we have to compute costly disconnected contributions requiring large numbers of propagator inversions to extract a signal. As the computational cost of these inversions scales like some power ($V^n,n>1$) of the volume and require more iterations as the quark mass decreases, a small volume was deemed a necessity to broadly and accurately map the spectrum for a large range of quark mass. Investigation of the finite volume effects from our volume and the finite lattice spacing effects will be left for a future study.

The bare gauge coupling $\beta=4/g^2$ is a function of the lattice spacing, which serves as the ultraviolet cutoff. For this study we choose to work at fixed bare gauge coupling of $\beta=2.2$.
The physical scale of our theory can be defined by matching to a known phenomenological scale.
The bare quark mass $m_0$ (which is typically a negative number) gets shifted by additive renormalization, so the massless limit can only be found from the results of numerical simulations.  We calculate with several different values of $m_0$ as listed in Table~\ref{tab:lat_par}.
Also shown in Table~\ref{tab:lat_par} are the number of configurations in each ensemble.  Ensembles were generated using the RHMC algorithm \cite{Clark:2006fx}.

After these ensembles have been generated, the largest remaining expense is the calculation of quark propagators, which requires inversion of a large-but-sparse matrix,
\begin{equation}
M(x,y) = (4+m_0)\delta_{x,y} - \frac{1}{2}\sum_{\mu=1}^4\bigg((1-\gamma_\mu)U_\mu(x)\delta_{x+\mu,y}+(1+\gamma_\mu)U_\mu^\dagger(x)\delta_{x-\mu,y}\bigg) \,.
\end{equation}
For many applications only one row of the inverse is required and then it is sufficient to solve the eigenvalue problem
\begin{equation}
M(x,y)S(y) = \eta(x)
\end{equation}
and we choose the source to be a time-diluted \cite{Bernardson:1993he} $Z_2$-stochastic wall (Z2SEMWall) \cite{Boyle:2008rh} source.
Unfortunately, one row is not sufficient whenever two quarks within a single operator can annihilate.
For these disconnected diagrams, we use an unbiased stochastic estimator, i.e. time and spin dilution \cite{Bitar:1988bb,Bali:2009hu}. We find that $64$ stochastic ``hits" per configuration was beneficial for reducing noise at reasonable cost, which is a finding similar to \cite{Arthur:2016ozw}. From the configurations listed in Tab.~\ref{tab:lat_par} for our lightest quark masses we do approximately $O(64,000\rightarrow 330,000)$ inversions for each ensemble's spectrum measurement.

\subsection{The lightest hadrons}
\label{sec:LightHadrons}

To create a hadron on a lattice, we select the appropriate operator from Eqs.~(\ref{eq:operators}). %apply it at a particular Euclidean time, and sum over all points on this time-slice to project zero momentum.  The conjugate operator is similarly applied at a different Euclidean time to annihilate the hadron, and the resulting correlation function will depend exponentially on the hadron mass.  More precisely, 
%% Yo, we don't "sum over spatial directions" we sum over the points on that time-slice, I find this terminology not at all clear. Also the dagger is the creation (source) operator. Plus all of this intro is too waffly. Pheno people will totally get this part as it is standard continuum field theory - J
We create the state at some initial euclidean time and destroy it at some different time, so the resulting correlation function is given by
\begin{equation}
\begin{aligned}
C_{\mathcal{O}_1 \mathcal{O}_2}(t) &= \frac{1}{L^3}\sum_x \langle \mathcal{O}_1(x,t) \mathcal{O}_2^\dagger(0,0) \rangle,\\
&= \sum_n \frac{\langle 0 | \mathcal{O}_1 | n \rangle \langle n | \mathcal{O}_2 | 0\rangle }{2m_n} \left( e^{-m_n t}\pm e^{-m_n(T-t)}\right).
\end{aligned}\label{eq:BasicCorr}
\end{equation}
Hadron masses $m_n$ can be obtained by fitting the lattice data to this functional form.
The calculation for our dark matter candidate, the $\rho^\pm$ of Eq.~(\ref{eq:rhomultiplet}), is straightforward as we can choose an operator where each quark propagator runs from source to sink, but other hadrons are much more costly due to the ability of $\bar qq$ to annihilate within a single operator. 

We also compute the decay rates of dark sector states. 
Notice that Eq.~(\ref{eq:BasicCorr}) contains $\langle0|{\cal O}_1|n\rangle$, which is proportional to the hadron's decay constant and can be extracted from the lattice data up to a multiplicative renormalization factor.
For this project, we compute 
\begin{equation}\label{eq:decays}
Z_A \langle 0 | \mathcal{O}_A^t(0) | \eta \rangle = f_\eta m_\eta,\quad
Z_V \langle 0 | \mathcal{O}_V^{ai}(0) | \rho^a \rangle = f_\rho m_\rho \hat{e}_i ,\quad
Z_P \langle 0 | \mathcal{O}_P(0) | \eta \rangle = f_P \frac{m_\eta^2}{m_q}\, ,
\end{equation}
where $i$ ($t$) is a spatial (temporal) Lorentz component and $\hat{e}$ is a unit polarization vector.
We find it beneficial to simultaneously fit
\begin{equation}\label{eq:combined_pseudo}
\begin{aligned}
C_{\mathcal{O}_A^t \mathcal{O}_P}(t) &= \sum_{n=1}^N A_n B_n \left( e^{-m_\eta^n t} - e^{-m_\eta^n (T-t)}\right),\\
C_{\mathcal{O}_P\mathcal{O}_P}(t) &= \sum_{n=1}^N B_n^2 \left( e^{-m_\eta^n t} + e^{-m_\eta^n (T-t)}\right),
\end{aligned}
\end{equation}
for the pseudoscalar to determine the ground state mass and amplitude, and
\begin{equation}\label{eq:combined_vector}
\begin{aligned}
C_{\mathcal{O}_T^{ti} \mathcal{O}_T^{ti}}(t) &= \sum_{n=1}^N C_n^2 \left( e^{-m_\rho^n t} + e^{-m_\rho^n (T-t)}\right),\\
C_{\mathcal{O}_V^i\mathcal{O}_V^i}(t) &= \sum_{n=1}^N D_n^2 \left( e^{-m_\rho^n t} + e^{-m_\rho^n (T-t)}\right),
\end{aligned}
\end{equation}
for the vector. We find that over the whole temporal range (excluding $t/a=0$) a multi-cosh/sinh fit with three states ($N=3$) does a good job of describing our data. $n=1$ is our lightest state, and so we can use the fit parameters in the following way to define the decay constants of Eq.~\eqref{eq:decays},
\begin{equation}\label{eq:f_computations}
\begin{aligned}
f_\eta = Z_A A_1\sqrt{\frac{2}{m_\eta^1}},\quad f_\rho = Z_V D_1 \sqrt{\frac{2}{m_\rho^1}},\quad f_P = Z_P B_1 \sqrt{\frac{2m_q}{(m_\eta^1)^2}} \\
\end{aligned}
\end{equation}
For the renormalisation factors we take the results from \cite{DelDebbio:2008wb} who determined them using 1-loop perturbation theory (see also \cite{Hietanen:2014xca}),
\begin{equation}
Z_{A/V/P} = 1 - \frac{g_0^2}{16\pi^2}\frac{3}{4}C_{A/V/P}
\end{equation}
with coefficients $C_A=15.7$, $C_V=20.62$, and $C_P=-6.71$.

In a theory with more than one quark flavour, there would be a pseudo-Goldstone boson like the pion of QCD for which lattice calculations do not require disconnected contributions and so can give a very precise determination of the mass. 
In our single flavour theory, this state is absent.
Nevertheless, we calculate this fictitious mass, which we denote as $m_\pi$, by neglecting disconnected contributions to Eq.~\eqref{eq:combined_pseudo}.
Even though $m_\pi$ is not a state in our theory, it provides a convenient alternative to the renormalized quark mass and allows us to determine the ``chiral'' point at which the quark mass vanishes.
%Fig.~\ref{fig:PionExtrap} of App.~\ref{app:mc} confirms that our calculations of $m_\pi^2$ are consistent with a linear dependence on the bare mass. -- There is no need for this statement, people can see it if they wish - J
The extrapolation to $m_\pi^2=0$ defines a critical value of bare quark mass $m_0$ that we will call $m_c$. Numerically, we find
\begin{equation}\label{eq:mc}
m_c = -0.9029(4).
\end{equation}
Another way to determine $m_c$ uses a calculation of the topological susceptibility and leads to a compatible result, as shown in App.~\ref{app:topo}.
We define the quark mass to be $m_q = m_0 - m_c$.
%% didn't like the "in our model" we are simulating a specific theory which we have a model to justify.

\subsection{String tension and confinement scale}\label{sec:sigmaLambda}

In lattice calculations, dimensionful quantities are given in units of the lattice spacing $a$ and must be determined by fixing a physical scale. 
Since there are no known fixed scales to normalize our dark sector model, we will use the the dark confinement scale $\LambdaMS$ to define the overall scale of our theory. We will then report masses and decay constants in units of $\LambdaMS$. As in QCD, $\LambdaMS$ is the characteristic energy scale of strong interactions. %% didn't like what was here before as it was far too verbose

While $\LambdaMS$ is purely defined in perturbation theory, we can perturbatively match it to the string tension $\sigma$, which can be directly measured in simulations. Following Eqs.~(4.60) and (4.61) of \cite{Bali:2000gf} and perturbative factors from \cite{Schroder:1998vy},
the result for our $\SU(2)$ theory with one fundamental quark flavour is
\begin{equation}\label{eq:sigma2Lambda}
\LambdaMS = 0.7712\sqrt{\sigma} \, .
\end{equation}

The string tension is the slope of the linear potential between two color charges at large separation. 
The potential between a static quark and static anti-quark can be measured by tracing over Wilson loops connecting points $x$ and $x+r$ which have temporal length $\tau$ (see \cite{Bali:2000gf} and references therein),
\begin{equation}
W(r,\tau) \underset{\tau\gg 0}{=} A e^{-aV(r)\tau}.
\end{equation}
Since generating these Wilson loops for each possible separation $r/a$ is somewhat expensive, following \cite{Bernard:2000gd}, we fix the fields to Coulomb gauge and then need only compute open-ended Polyakov line correlators because the gauge condition will connect the ends of the line spatially. We start by measuring all matrix-valued Polyakov lines $P$, of length $\tau$, from timeslice $T$, over $L^3$, as
\begin{equation}
P(x,\tau) = \prod_{t=T}^{\tau+T} U_{\hat{t}}(x,t).
\end{equation}
We can then directly compute the quantity 
\begin{equation}
W(r,\tau) = \text{Tr}\left[ P(x,\tau) P^{\dagger}(x+r,\tau)\right],
\end{equation}
for all separations $r$ (and all of their translations over the $L^3$ volume) cheaply by performing the convolution with fast Fourier transforms,
\begin{equation}
\begin{gathered}
W(r,t) = \frac{1}{L^3}\sum_q e^{-iq\cdot r} \text{Tr}\left[ P(q,\tau) P^\dagger(q,\tau)\right],\quad P(q,\tau) = \sum_x e^{iq\cdot x} P(x,\tau).
\end{gathered}
\end{equation}
We then repeat this operation over all possible timeslices to improve the statistics of this still-quite-noisy quantity. An important feature of this definition is that it manifestly incorporates the periodicity of the gauge fields and so will correctly average loops with one line at $x$ and the other at $x+1$ or $x+L-1$. The largest separation we can have in any one direction is therefore $L/2$. We will average over equivalent $r^2$ values to further boost statistical precision.

We can investigate where the static potential has saturated its ground state by looking at an ``effective mass'',
\begin{equation}\label{eq:statpot_mass}
V(r) = -\log\left(\frac{W(r,\tau+1)}{W(r,\tau)}\right).
\end{equation}
The signal degrades for large values of $\tau$ and suffers from excited state contamination at very small $\tau$, so an appropriate middle-ground must be taken.
The static potential is often fit to the Cornell-type model \cite{Eichten:1974af},
\begin{equation}\label{eq:cornell}
V(r) = \frac{A}{r} + B + \sigma \;r \,.
\end{equation}
%where the string tension $\sigma$ has units of energy/length. - isnt this obvious?
The dimensionless quantity that is extracted from the lattice simulation is $a^2\sigma$. %, where $a$ denotes the lattice spacing. - repetition again!!
%Therefore phenomenological input for the value of $\sigma$ determines the lattice spacing, which then allows the conversion of lattice predictions for masses and decay constants into physical units.

Based on our calculation of the massless limit in Eq.~(\ref{eq:mc}), we  fit simultaneously the mass dependence of Eq.~(\ref{eq:cornell}) with
\begin{equation}\label{eq:statpot_massdep}
V(r) = B(1+c_1m_q) + a^2\sigma(1+c_2m_q)r.
\end{equation}
For the string tension, we only need to fit the constant and linear terms of the Cornell potential. The fit parameters $B$ and $a^2\sigma$ then give the potential in the massless-quark limit.
We note that at small $r$ there are significant discretisation effects, and at large $r$ we expect significant signal deterioration and finite volume effects. Hence, we have performed our fits between these extremes, performing a fit-window analysis where we varied the upper and lower ends of the window looking for both a minimum in $\chi^2$ and stability in the fit parameter $a^2\sigma$.
We then used a representative fit window to obtain our quoted results.

\begin{figure}[t!]
\centering
{
\includegraphics[scale=0.275]{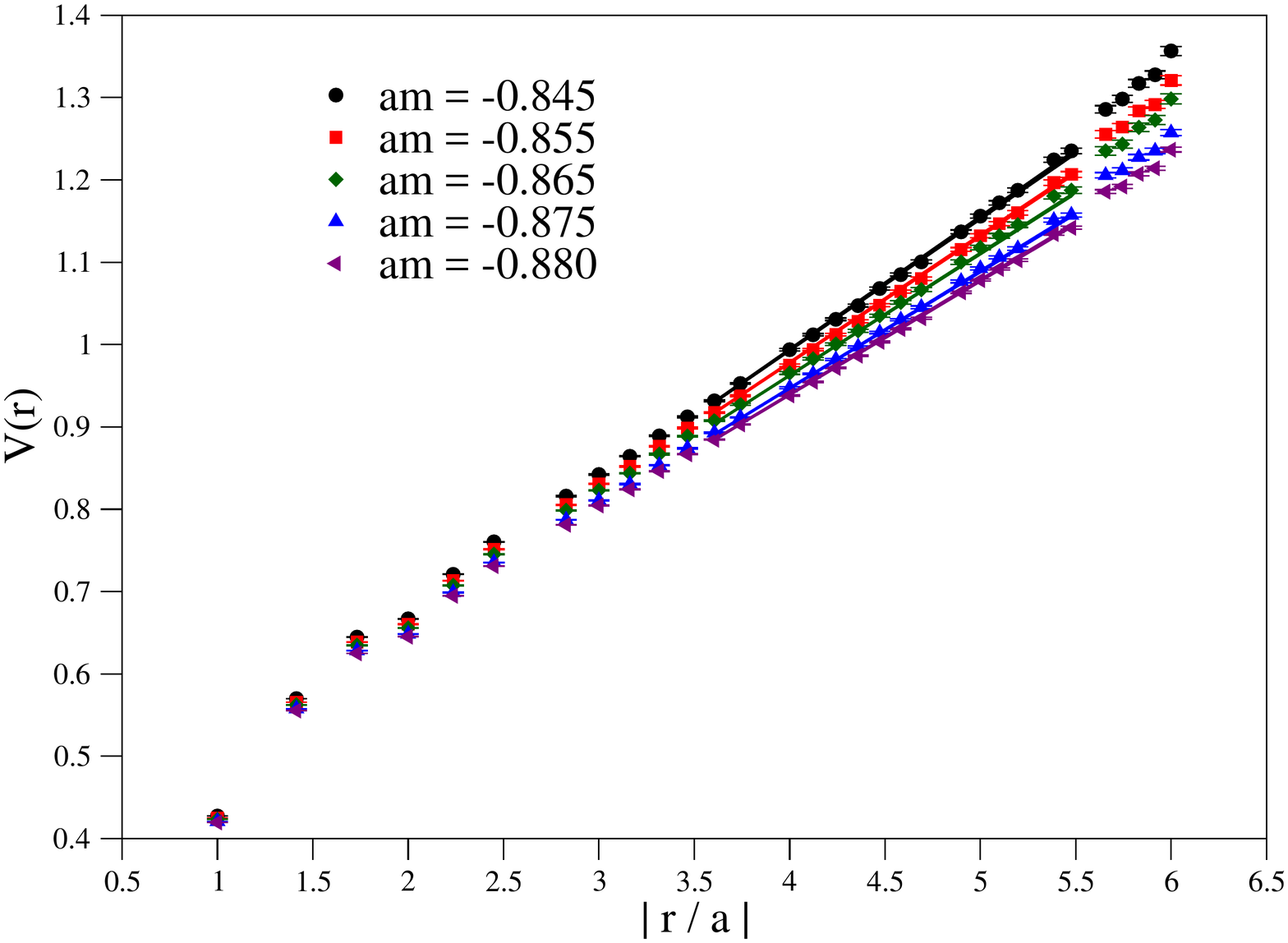}
}
\hspace{2pt}
{
\includegraphics[scale=0.275]{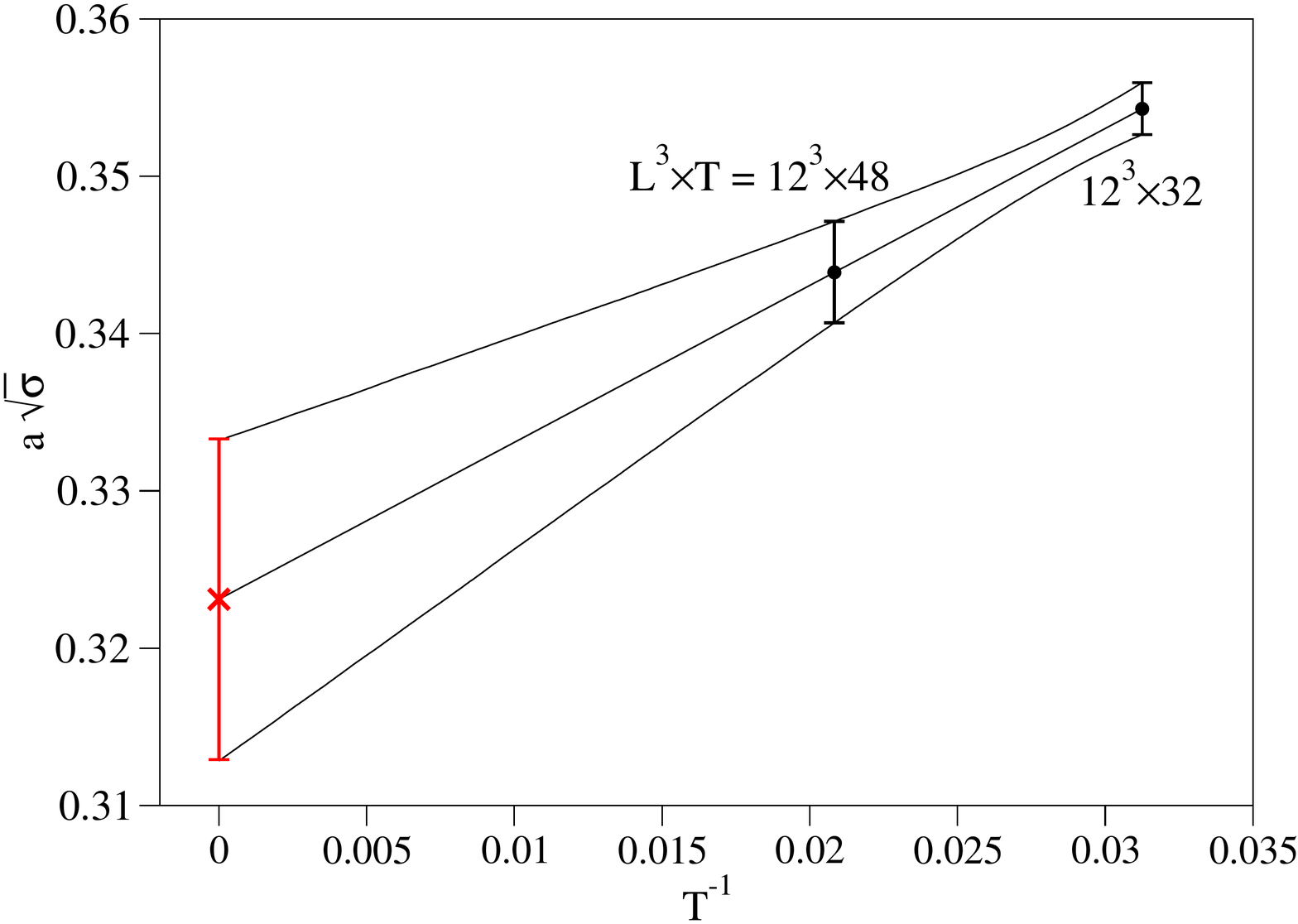}
}
\caption{Left panel: Fit of the $12^3\times32$ static potential to Eq.~(\ref{eq:statpot_massdep}).
Right panel: Extrapolation of the square root of the string tension.}\label{fig:statpot_fit}
\end{figure}

\begin{table}[b!]
\caption{Global fit results for the static potential with $T=32$. We used $\tau=2$ in Eq.~(\ref{eq:statpot_mass}) to determine $aV(r)$.}\label{tab:statpot_params}
\begin{tabular}{c|cc|cc|cc|c}
\toprule
$T$ & $B$ & $c_1$ & $a^2\sigma$ & $c_2$ & $\chi^2$/dof \\
\hline
32 & 0.405(4) & -2.5(3) & 0.1255(12) & 5.0(3) & 2.6 \\
48 & 0.429(7) & -2.7(4) & 0.1183(22) & 5.9(6) & 0.7 \\
\botrule
\end{tabular}
\end{table}

Fits for our $12^3\times32$ lattices are displayed in the left panel of Fig.~\ref{fig:statpot_fit} and numerical results for both volumes are listed in Tab.~\ref{tab:statpot_params}.
Our final result for the string tension is
\begin{equation}
a\sqrt{\sigma} = 0.323(10)\,,
\end{equation}
and that combines with Eq.~(\ref{eq:sigma2Lambda}) to give
\begin{equation}
a\LambdaMS = 0.249(8) \,.
\end{equation}
This auxilliary scale allows us to quote all of our results in terms of the physical confinement scale that phenomenologists can choose, $\LambdaMS$, instead of the dimensionless quantities we directly compute.
%Whenever the desired value of $\LambdaMS$ is given from phenomenology, this result can be used to determine the lattice spacing $a$ and from that the scale of all dimensionful lattice quantities. -- I REALLY didn't like this as it suggests that pheno people should actually measure our lattice spacing, instead of just dialling the physical Lambda to regions they like. There is no measurement of this scale (yet or perhaps ever) so it is a choice they make - J
% In the following sections, we will plot results scaled by $\LambdaMS$ rather than by the lattice spacing. -- I don't like this either, too much repetition

When high precision is required, lattice QCD studies typically set the scale with quantities called $t_0$ and $w_0$ rather than using the string tension. Appendix~\ref{sec:t0w0} presents our calculation of those quantities.
However, for our exploratory study, $\LambdaMS$ is convenient, perhaps more phenomenoligically relevant, and entirely sufficient.

%% hadron masses and decay constants
\section{Hadron masses and decay constants}
\label{sec:spectrum}

\begin{figure}[h!]
\centering
\includegraphics[width=0.7\textwidth]{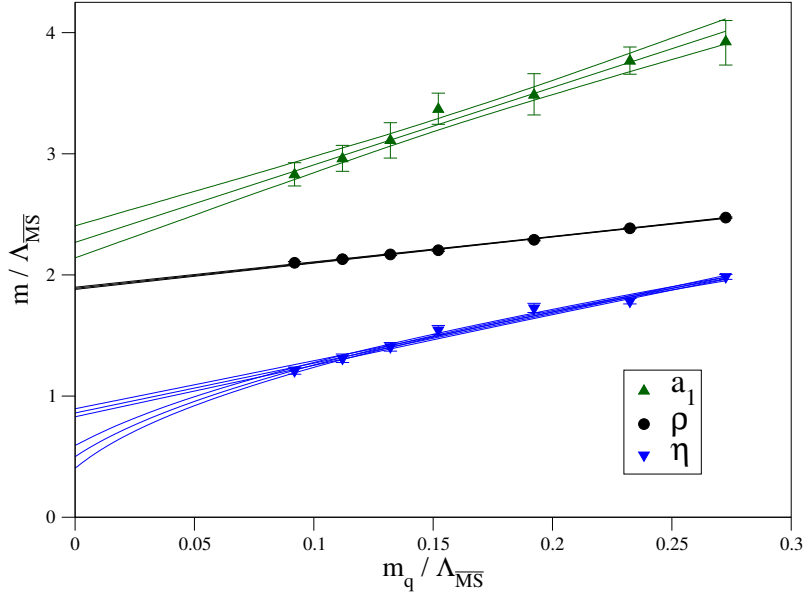}
\caption{Masses of the $\eta$, $\rho$, and $a_1$ hadrons for our five lightest quark masses and the fits used to determine their values in the massless quark limit.}
\label{fig:mlight}
\end{figure}

Fig.~\ref{fig:mlight} shows the masses of some of our lightest hadrons from simulations with our lightest bare quark masses: the pseudoscalar $\eta$, vector $\rho$ and the axial vector hadron $a_1$. The $\eta$ appears about a factor of 2 lighter than the $\rho$ and a factor of 3 lighter than the $a_1$ for our lightest simulated quark mass. The $a_1$ is noticeably heavier than the $\rho$ but approaches it in our massless-quark limit.  Disconnected diagrams have been omitted from the axial vector calculation because they were found to be too noisy to make any quantifiable contribution. 

All three hadrons' mass dependence for small quark masses can be described quite well by a linear fit in $m_q$, as is illustrated in Fig.~\ref{fig:mlight}. However, the mass-dependence of the $\eta$ appears to slightly prefer the form $m_\eta^2=a+bx$, giving a $\chidof\approx1.5$. This is in line with the na\"ive expectation that the $\eta$ receives a constant shift due to the anomaly even at vanishing quark mass \cite{Creutz:2006ts}.

It is worth noting that the determination of $m_\eta$ for $m_q/\LambdaMS=0.233$ is particularly low, and this is probably the reason for poor fits at larger quark mass. We have also noticed that the disconnected contribution to the $\eta$ becomes more difficult to measure at larger quark masses.

\begin{table}[h!]
\centering
\begin{tabular}{c|ccc|ccc}
\toprule
~$m_q/\Lambda_{\overline{\rm MS}}$~ & ~$m_\eta/\Lambda_{\overline{\rm MS}}$~ & ~$m_\rho/\Lambda_{\overline{\rm MS}}$~ & ~$m_{a1}/\Lambda_{\overline{\rm MS}}$~ & ~$f_\eta/\Lambda_{\overline{\rm MS}}$~ & ~$f_\rho/\Lambda_{\overline{\rm MS}}$~ & ~$f_P/\Lambda_{\overline{\rm MS}}$~\\
\hline
%% previous results
% 0.093 & 1.178(9)  & 2.110(7)  & 2.93(11) & 0.137(3) & 0.582(3) \\
% 0.113 & 1.261(14) & 2.129(9)  & 2.95(9)  & 0.155(4) & 0.594(3) \\
% 0.133 & 1.424(7)  & 2.146(13) & 3.10(16) & 0.173(5) & 0.597(5) \\
% 0.153 & 1.489(17) & 2.206(10) & 3.42(17) & 0.178(5) & 0.607(4) \\
% 0.193 & 1.695(13) & 2.288(9)  & 3.56(14) & 0.186(5) & 0.624(4) \\
% 0.233 & 1.732(13) & 2.388(5)  & 3.80(20) & 0.200(3) & 0.640(2) \\
% 0.274 & 1.990(15) & 2.475(5)  & 4.02(10) & 0.212(4) & 0.648(2) \\
0 & 0.500(94)/0.861(33) & 1.889(9) & 2.27(13) & 0.078(18) & 0.628(16) & 0.364(5) \\
\hline
0.092 & 1.209(29) & 2.110(11) & 2.83(10) & 0.157(4) & 0.709(10) & 0.385(4)\\
0.112 & 1.311(35) & 2.130(8) & 2.96(11) & 0.177(6) & 0.732(3) & 0.393(5) \\
0.132 & 1.407(35) & 2.169(9) & 3.11(15) & 0.199(8) & 0.752(6) & 0.401(7)\\
0.152 & 1.544(37) & 2.204(9) & 3.37(13) & 0.205(7) & 0.760(4) & 0.405(6)\\
0.192 & 1.726(38) & 2.290(8) & 3.48(17) & 0.216(7) & 0.777(4) & 0.425(5)\\
0.234 & 1.783(23) & 2.385(5) & 3.77(11) & 0.226(4) & 0.794(2) & 0.417(4)\\
0.273 & 1.985(20) & 2.473(5) & 3.92(18) & 0.241(4) & 0.803(2) & 0.436(3) \\
\hline
$\chidof$ & 1.5/2.0 & 0.9 & 0.3 & 1.4 & 0.4 & 1.1 \\
\botrule
\end{tabular}
\caption{Numerical values for some masses and decay constants from our lightest quark simulations. Values at zero quark mass are obtained from various extrapolations as explained in the text.}
\label{tab:outputs}
\end{table}

%%%%%%%%%%%%%%%%%%%%%%%%%%%%%%%%%%%%%%%%%%%%%%%%%%%%%%%%%%%%%%%%
%%%%%%%%%%%%%%%%%%%%%%%%%%%%%%%%%%%%%%%%%%%%%%%%%%%%%%%%%%%%%%%%

%%%%%%%%%%%%%%%%%%%%%%%%%%%%%%%%%%%%%%%%%%%%%%%%%%%%%%%%%%%%%%%%
\begin{figure}[h!]
\centering
\includegraphics[width=0.7\textwidth]{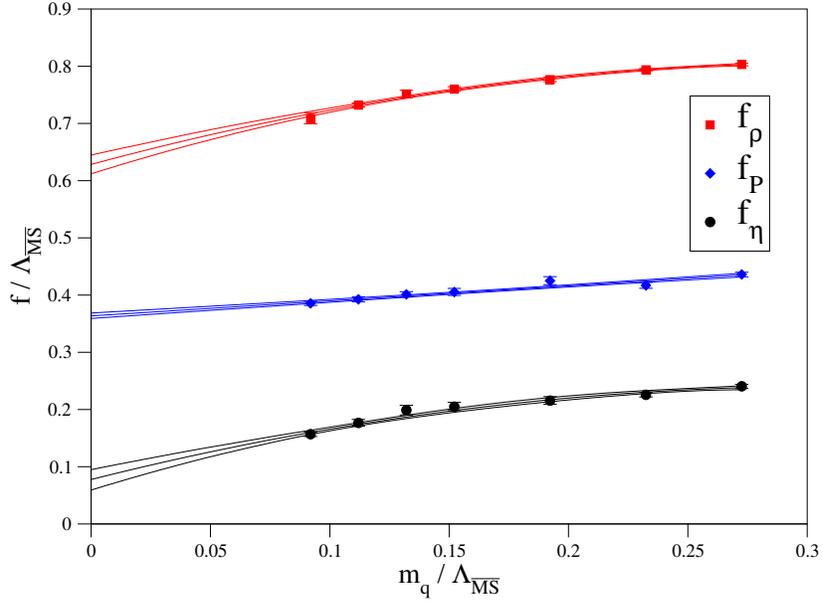}
\caption{The quantities $f_\eta$, $f_\rho$, and $f_P$ for our lightest masses. With the fits used to determine their results in the massless limit.
}
\label{fig:DecayConstants}
\end{figure}
%%%%%%%%%%%%%%%%%%%%%%%%%%%%%%%%%%%%%%%%%%%%%%%%%%%%%%%%%%%%%%%%

The decay constants $f_\eta$, $f_\rho$, and $f_P$ of Eq.~\eqref{eq:decays} are displayed in Fig.~\ref{fig:DecayConstants}.
$f_\rho$ is approximately $\frac{3}{4}\LambdaMS$ and $f_\eta$ is much smaller, ranging from $\frac{1}{3}$ to $\frac{1}{5}$ the size of the $\rho$ decay constant over the handful of masses in Tab.~\ref{tab:outputs}. The decay constant $f_P$ is of importance to DM phenomenology (c.f~\ref{sec:fate_DM}) and we find that its value is roughly twice the size of the $\eta$ decay constant over the quark mass range considered here, and shows little sign of curvature.

We found that a simple linear fit describes the data of $f_P / \LambdaMS$ very well over our range of lightest quark masses and this is evident in the plot. However, some level of curvature appears present in both $f_\rho$ and $f_\eta$. We believe this to be the onset of higher-order corrections of $m_q$ or some other functional dependence affecting our extrapolation and so find the data to be reasonably well described by the quadratic form $f_{\rho/\eta}=a+bm_q+cm_q^2$.

\begin{figure}
\centering
\includegraphics[width=0.8\textwidth]{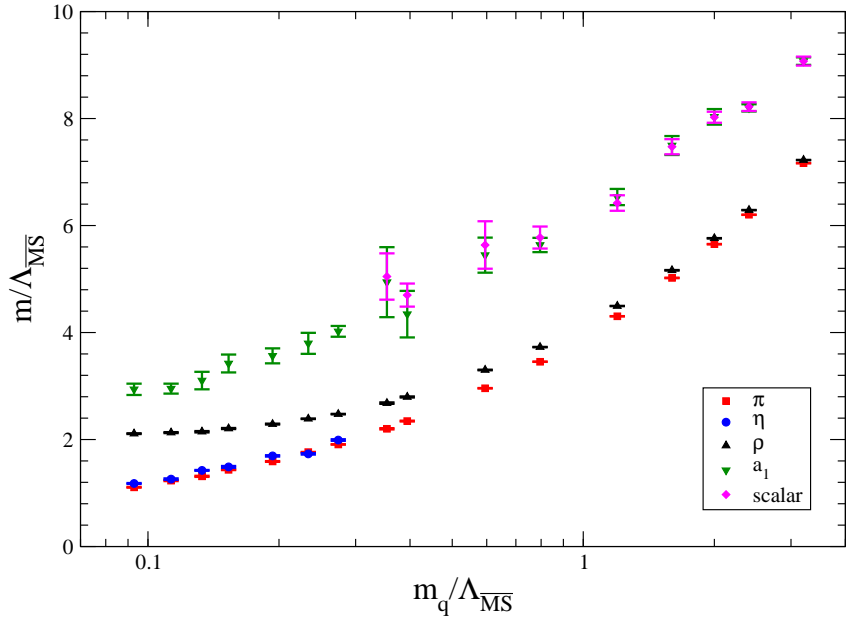}
\caption{Hadron masses for the simulation parameters listed in Tab.~\ref{tab:lat_par}.}
\label{fig:mheavy}
\end{figure}

An overview of the broad mass spectrum in this minimal dark theory is given by Fig.~\ref{fig:mheavy}.
%% Put this all into perspective - compare to a known strongly-interacting theory QCD
It appears as though this theory is somewhat reminiscent of QCD in the heirarchy of its spectrum; it has a light pseudoscalar meson, a heavier vector meson, and heavier still axial and scalar mesons. The decay constant for the pseudoscalar is smaller than that of the vector by about a factor 4, which in QCD is about a factor 2 different. Our dark matter candidate sits at roughly $2\LambdaMS$, which is in the same ballpark as the $\rho$-meson in QCD.
With dark matter phenomenology in mind, we note that $m_\eta<m_\rho$ appears true for any value of the quark mass.

The extension of Fig.~\ref{fig:mheavy} to larger $m_q$, however, should be viewed with caution. 
Lattice artifacts can become large where $am_q > 1$.
Nevertheless, the right side of Fig.~\ref{fig:mheavy} shows a phenomenon familiar from heavy quark physics: hyperfine splittings shrink to produce a degeneracy of pseudoscalar with vector and also scalar with axial vector.

%%%%%%%%%%%%%%%%%%%%%%%%%%%%%%%%%%%%%%%%%%%%%%%%%%%%%%%%%%%%%%%%
%%%%%%%%%%%%%%%%%%%%%%%%%%%%%%%%%%%%%%%%%%%%%%%%%%%%%%%%%%%%%%%%

%% Higgs coupling (i.e. the sigma term)
\section{Dark sector phenomenology}
\label{sec:sigmaterm}

\subsection{Direct detection and Higgs decay}

The lightest vector meson is our DM candidate and is represented by an iso-triplet vector field $\rho_\mu^a$, where $a$ labels baryonic isospin.
At lowest order, there are two operators, $\mathcal{O}_{S,P}$, that may couple to the Higgs field, as given in Eq.~\eqref{eq:higherdim}.
Since the CP phase $\phi$ is arbitrary, we may treat the corresponding mass scales $M_S = M/\cos\phi$ and $M_P = M/\sin\phi$ as separate parameters.

Including the scalar operator, the low-energy effective Lagrangian for DM is
\begin{equation} \label{eq:Leff}
\mathcal{L}_{\rm eff} \supset - \frac{1}{4} \rho^a_{\mu\nu} \rho^{a \mu\nu} + \frac{1}{2} m_\rho^2 \, \rho^a_\mu \rho^{a \mu}
- \frac{1}{2} \lambda_S \, \rho^a_\mu \rho^{a\mu} \left(|H|^2 - \tfrac{1}{2}v^2\right)
\end{equation}
where $\rho_{\mu\nu}^a$ is the field strength tensor and $\lambda_S = \langle \rho|\bar{q} q |\rho \rangle /M_S$ is a coupling determined below.\footnote{We expect the pseudoscalar term in Eq.~\eqref{eq:higherdim} to induce a Higgs-DM interaction of the form $\varepsilon^{\alpha\beta\mu\nu} \rho_{\alpha\beta}^a \rho_{\mu\nu}^a (|H|^2 - \tfrac{1}{2} v^2) /M_P$. However, this leads to a velocity-suppressed direct detection cross section that is much less constrained compared to the scalar interaction.} 
We have omitted purely dark sector interactions, e.g., with the $\eta$ meson, that are beyond the scope of this work.

The low-energy theory of dark baryons in Eq.~\eqref{eq:Leff} is reminiscent of models of hidden vector DM coupled via the Higgs portal~\cite{Hambye:2008bq,Kanemura:2010sh,Djouadi:2011aa,Lebedev:2011iq,Baek:2012se}. 
In these models, one typically assumes that the Higgs interaction governs the DM relic abundance, implying a lower bound on $\lambda_S$. 
This parameter space is strongly constrained by a combination of direct detection and Higgs decay limits~\cite{Djouadi:2011aa}.
In our framework, however, this assumption is not necessary since strong dynamics within the dark sector determine the relic density.

\begin{figure}
\centering
\includegraphics[width=0.7\textwidth]{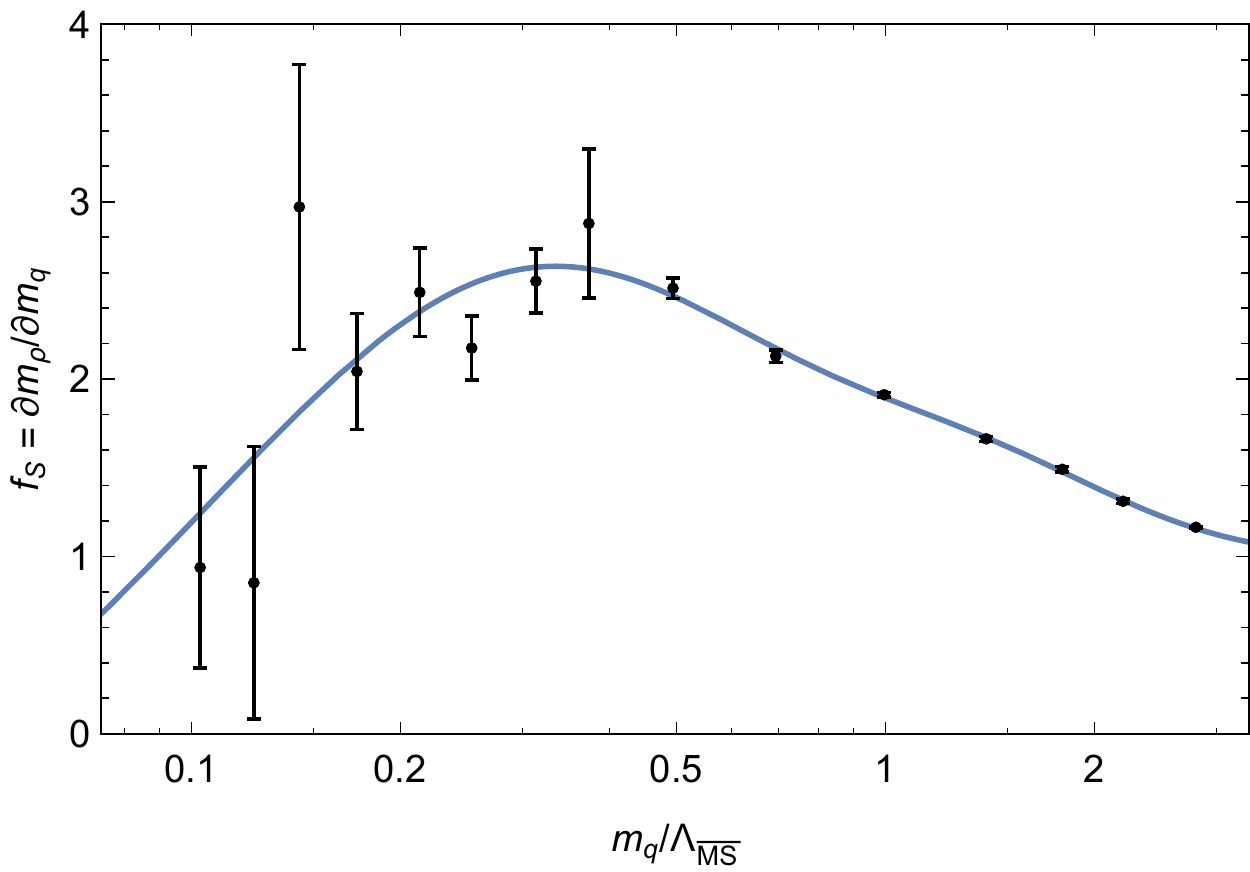}
\caption{Scalar form factor $f_S$ entering into the Higgs-DM coupling, obtained from finite differences (points) and from an analytic fit  (solid curve). The analytic fit is  $m_\rho/\LambdaMS = \mathcal{F}(m_q/\LambdaMS)$ where
$\mathcal{F}(x) = (2.12 + 0.87\,x + 10.60 \,x^2) e^{-3.94\, x} + (4.10 + 1.00\,x)(1 - e^{-1.42\, x})$ and has a $\chi^2/\textrm{dof} \approx 1.1$.}
\label{fig:sigmaterm}
\end{figure}

The spin-independent DM-nucleon cross section is~\cite{Kanemura:2010sh}
\begin{equation}
\sigma_{\rho N} = \frac{\lambda_S^2 m_N^4 f_N^2}{4 \pi m_h^4 (m_\rho + m_N)^2} \, ,
\end{equation}
where $m_h$ is the Higgs boson mass and $f_N \approx 0.3$ is the Higgs-nucleon coupling~\cite{Hoferichter:2017olk}.
The coupling $\lambda_S$ depends on a matrix element determined by our lattice results.
The Feynman-Hellman theorem allows us to write
\begin{equation} \label{eq:scalar}
\langle \rho | \bar{q} q |\rho \rangle = \frac{\partial m_\rho^2}{\partial m_q} = 2 m_\rho f_S\, ,
\end{equation}
where $f_S = \partial m_\rho/\partial m_q$. 
We determine $f_S$ from our lattice results using two methods.
First, we perform an analytic fit to the data points for $m_\rho$ in Fig.~\ref{fig:mheavy} and take the derivative.
Second, we compute the derivative using the finite differences of the points.
Both methods, shown in Fig.~\ref{fig:sigmaterm}, are in good agreement and yield values in the range $f_S\approx 1-3$.
However, lattice artifacts are present for $m_q/\LambdaMS \gtrsim 1$ (corresponding to $m_\rho \gtrsim a^{-1}$), likely leading to an underestimate of $f_S$.
We expect $f_S \approx 2$ at large quark mass since $m_\rho \approx 2 m_q$.
We additionally caution against extrapolating Fig.~\ref{fig:sigmaterm} to $m_q = 0$ since the intercept may vary wildly according to the function used to fit the data.

%%%%%%%%%%%%%%%%%%%%%%%%%%%%%%%%%%%%%%%%%%%%%%%%%%%%%%%%%%%%%%%%
%%%%%%%%%%%%%%%%%%%%%%%%%%%%%%%%%%%%%%%%%%%%%%%%%%%%%%%%%%%%%%%%

Direct detection limits are most constraining for weak-scale DM mass. Recently, XENON1T obtained the most stringent upper bound on the spin-independent cross section, $4.1 \times 10^{-47} \; \textrm{cm}^2$ for 30 GeV DM mass~\cite{Aprile:2018dbl}, implying $M_S > 28 \; \textrm{TeV}$. For larger DM mass, the XENON1T bound weakens while $\sigma_{\rho N}$ is nearly constant.

Higgs studies at the LHC provide the most stringent constraints for low mass DM. 
In our model, the Higgs boson may decay into dark sector states that are long-lived and escape the detector.
If we assume $m_q, \LambdaMS \ll m_h/2$ and that all dark states escape invisibly, it is straightforward to compute the Higgs invisible width from a quark-level calculation.
We have
\begin{equation} \label{eq:higgsinv}
\Gamma(h \to \textrm{inv}) = \frac{m_h v^2}{4 \pi M^2} \, ,
\end{equation}
which is independent of $m_\rho$, the CP phase $\phi$, or any other dark sector parameters.
Present limits constrain the Higgs invisible branching fraction to be below 23\%~\cite{Aad:2015pla,Khachatryan:2016whc}.
For our model, this implies $M > 40$ TeV.

We note that the invisible Higgs constraints are very different compared to hidden vector DM models where DM is a gauge boson, not a composite state.
In that case, the Higgs invisible width scales as $\Gamma(h \to \textrm{inv}) \propto m_{\rm DM}^{-4}$~\cite{Lebedev:2011iq} and becomes very constraining for light DM (see, e.g., Fig.~9 of \cite{Aad:2015pla}).

\subsection{Fate of the lightest dark hadron}\label{sec:fate_DM}

The lightest state in the dark spectrum is the $\eta$ meson. 
If it were stable, it would constitute an $\mathcal{O}(1)$ fraction of the DM density.
However, the $\eta$ meson is not a worthy DM candidate since it can mix with the Higgs boson through a dimension-five operator $\mathcal{O}_P |H|^2$, inducing it to decay to the SM.
Even if this operator is suppressed by the Planck scale, the $\eta$ lifetime would be much shorter than the age of the Universe.
Its decay products, moreover, are fixed by the SM Higgs couplings.

The lifetimes of meta-stable dark states are strongly constrained if they decay into visible SM particles.
Cosmic microwave background measurements exclude an $\mathcal{O}(1)$ fraction of meta-stable DM unless it decays prior to recombination, before $\sim 10^{13}$ s~\cite{Slatyer:2016qyl}. 
Decays occurring between $\sim 0.1 - 10^{12}$ s affect primodial abundances of light nuclei~\cite{Jedamzik:2006xz}.
In particular, decays via Higgs mixing are largely constrained to occur before $\sim 0.1$ s, otherwise the injection of hadrons into the plasma alters the neutron/proton ratio after weak interactions have frozen out~\cite{Fradette:2017sdd}.
However, the limits depend on the cosmological abundance of $\eta$ mesons before they decay, which we defer to future work.
Here, to be conservative, we require the lifetime to be $\tau_\eta < 1$ s.

The total $\eta$ width can be written as
\begin{equation} \label{eq:etadecayrate}
\Gamma_\eta = \tau^{-1}_\eta = \sin^2 \theta_{h\eta} \Gamma_h(m_\eta) + \Gamma(\eta \to hh) 
\, .
\end{equation}
The first term represents $\eta$ decays through Higgs mixing, where the mixing angle $\theta_{h\eta}$ is defined by
\begin{equation}
\tan 2\theta_{h \eta} 
= \frac{2v \langle 0| \mathcal{O}_P |\eta\rangle}{M_P (m_h^2 - m_\eta^2)} 
\, , 
\end{equation}
and $\Gamma_h$ is the total SM Higgs width (evaluated at $m_\eta$, not $m_h$). 
We have adapted results from Ref.~\cite{Bezrukov:2009yw} to get $\Gamma_h$ as a function of mass below bottom threshold, while for larger mass we take results from Ref.~\cite{Djouadi:1997yw}.
The second term in Eq.~\eqref{eq:etadecayrate} is an additional decay channel that opens for $m_\eta > 2m_h$. 

The combination of $\tau_\eta < 1$ s and invisible Higgs decay yields a lower limit $m_\eta > 228$ MeV and $m_\rho > 320$ MeV for the range of $m_q$ in Table~\ref{tab:outputs}. 
This conclusion is further bolstered by astrophysical constraints on self-interactions, discussed below.

%%%%%%%%%%%%%%%%%%%%%%%%%%%%%%%%%%%%%%%%%%%%%%%%%%%%%%%%%%%%%%%%
\begin{figure}
\centering
\includegraphics[width=0.7\textwidth]{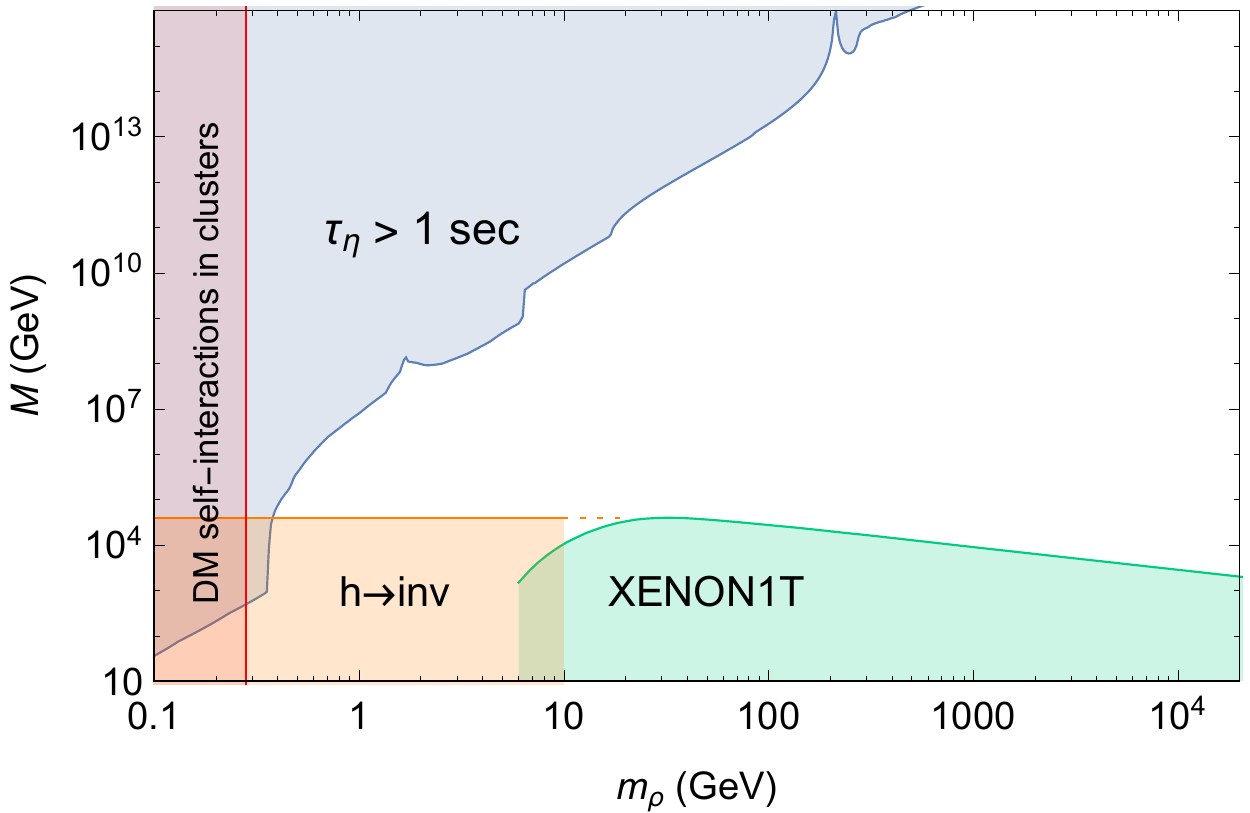}
\caption{Experimental and astrophysical constraints on our model, as a function of DM mass $m_\rho$ and the scale $M$ governing interactions between DM and the Higgs field. Shaded regions are excluded. This plot assumes $\phi = \pi/4$ and $m_q/\LambdaMS = 0.1$.}
\label{fig:pheno}
\end{figure}
%%%%%%%%%%%%%%%%%%%%%%%%%%%%%%%%%%%%%%%%%%%%%%%%%%%%%%%%%%%%%%%%

Fig.~\ref{fig:pheno} illustrates the complementarity between different constraints. 
For definiteness, we have taken $m_q/\LambdaMS = 0.1$ and CP phase $\phi=\pi/4$. 
The remaining parameters of the model are the DM mass $m_\rho$ and the interaction scale $M$. 
Other parameters of the model are determined according to our lattice results: $m_\eta \approx 0.57 \, m_\rho$, $f_P \approx 0.39$, and $f_S \approx 1$.
We have truncated the invisible Higgs limits at 10 GeV since the assumptions leading to Eq.~\eqref{eq:higgsinv} eventually breakdown.
With the exception of tuning $\phi=0$, taking other parameter choices does not greatly shift the shaded regions.

\subsection{Self-interactions}

In our model, DM particles are not collisionless and elastically scatter with one another through strong interactions.
If the scattering rate is large enough, self-interactions can leave an observable imprint on DM halos of galaxies and clusters.
The relevant figure of merit is $\sigma_\textrm{el}/m$, the cross section for DM elastic scattering per unit DM mass, which is typically expressed in units of $\textrm{cm}^2/\textrm{g} \approx 2 \; \textrm{barn}/\textrm{GeV}$.
While self-interacting DM is often motivated in terms of explaining various small scale structure issues~\cite{Tulin:2017ara}, here we simply make a conservative constraint on the parameter space of our model.
Actually calculating $\sigma_\textrm{el}/m$ is a challenging prospect for the lattice that we defer to future work.

By dimensional analysis, we expect $\sigma_\textrm{el} \sim 4 \pi \LambdaMS^{-2}$ since $\LambdaMS$ sets the typical size of $\rho$.
Since $m_\rho > 2 \LambdaMS$ for any dark quark mass, we can therefore set a lower bound
\begin{equation} \label{eq:sigmam}
\sigma_\textrm{el}/m \gtrsim 16 \pi/m_\rho^3 \, .
\end{equation}
Observations of relaxed massive clusters~\cite{Newman:2012nv,Newman:2012nw} provide the strongest constraint on self-interactions, favoring $\sigma_\textrm{el}/m \approx 0.1 \; \textrm{cm}^2/\textrm{g}$ or less~\cite{Kaplinghat:2015aga}.
If we take $\sigma_\textrm{el}/m < 0.5 \; \textrm{cm}^2/\textrm{g}$ as a conservative upper limit~\cite{Elbert:2016dbb}, we have
\begin{equation} \label{eq:si_limit}
m_\rho > 280 \; \textrm{MeV} \, .
\end{equation}
Merging cluster constraints, such as the Bullet Cluster~\cite{Randall:2007ph}, are comparatively weaker. 
In particular, recent simulations have found offsets for self-interacting DM halos to be much smaller than previously thought~\cite{Robertson:2016xjh}.

Our dimensional analysis estimate breaks down if DM scattering has an $s$-wave resonance, corresponding to a di-baryon $(\rho\rho)$ that is a nearly zero energy bound state.
In this case, $\sigma_{\textrm{el}}/m$ can be far larger than the lower bound implied by Eq.~\eqref{eq:sigmam}, approaching the $s$-wave unitarity limit when the mass gap and scattering energy go to zero~\cite{Braaten:2013tza}.
This is analogous to proton-neutron scattering, which is enhanced owing to the smallness of the deuteron binding energy.
Eq.~\eqref{eq:si_limit} is still satisfied in this case.
On the other hand, antiresonances (the Ramsauer-Townsend effect) may act to suppress DM scattering for certain choices of parameters~\cite{Tulin:2012wi,Tulin:2013teo}, evading our limit, but without a detailed calculation it is not possible to say anything further.

%% conclusions
\section{Conclusions}
\label{sec:conclusions}

Since strong dynamics explains the mass and stability of visible baryons, it is possible that similar physics is realized for DM as well.
In this work, we have studied the simplest model of dark baryons: $\SU(2)$ gauge theory with one flavor of dark quark.
Unlike QCD, the theory has no spontaneously broken chiral symmetries and no pseudo-Goldstone bosons.
Instead, there is an unbroken global $\SU(2)_B$ baryon symmetry resembling isospin, which unifies baryons and mesons into degenerate iso-multiplets.
The lightest baryon is one component of a iso-triplet vector $\rho$, which is our DM candidate.
Dark hadrons may couple to the SM through non-renormalizable interactions and we have considered the leading dimension-five operators involving the Higgs field.

In this initial and exploratory study, we have used lattice simulations to compute the spectrum of the lightest dark hadrons.
The overall mass scale of the theory is unknown {\it a priori}. 
Hence, with an eye towards phenomenology, we have presented all dimensionful parameters normalized with respect to the confinement scale $\LambdaMS$ (computed from the string tension $\sigma$).
The dark quark mass $m_q$ is a free parameter and our simulations focus on the quark mass regime with $m_q/\LambdaMS \approx 0.1 \rightarrow 1$.
In this range, the lightest hadron is the iso-singlet pseudoscalar meson $\eta$.
We have included the effect of disconnected diagrams, which causes the $\eta$ to remain massive according to our extrapolation to $m_q=0$, as expected from the $\U(1)_A$ anomaly.
The iso-triplet vector $\rho$ is the next-to-lightest state. 
We have also presented results for the lightest axial vector and scalar, which remain heavier still.

We note that several sources of systematic error have not been accounted for.
As our volume is quite small, we expect significant finite volume effects, particularly for light quark masses. 
We also expect finite lattice-spacing artifacts to be present since our lattice spacing is somewhat coarse and our action is correct only up to $\mathcal{O}(a)$ discretisation effects.
However, for a first study, the broad brush strokes of this theory are what is important and we anticipate our results to be accurate at around the 10\% level with these systematics in mind. 
Now that we better understand the parameter space and the model's feasibility as a DM candidate, dedicated finite volume and continuum limit studies beyond fixed $L$ and $\beta$ will be necessary to refine our numerical predictions.

In our opinion, there are three nice features of our model worth re-emphasizing, apart from its minimality.
\begin{itemize}
\item {\it DM stability:} The accidental $\SU(2)_B$ baryon number symmetry is preserved up through operators of dimension-five.
From an effective theory point of view, our DM candidate is as stable as the proton (and a counterexample to arguments in Ref.~\cite{Appelquist:2015yfa}).
\item {\it CP violation and $\eta$ decay:} 
Including dimension-five operators, the dark quark receives a mass contribution from the Higgs field in addition to its bare mass.
Since both terms need not be aligned in general, there appears a CP phase that mixes the $\eta$ with the Higgs boson, allowing the $\eta$ to decay rapidly in the early Universe before nucleosynthesis.
\item {\it Annihilation channel:} 
Our model has a built-in mechanism for efficient annihilation to set the DM relic density, $\rho \rho \to \eta \eta$, with the $\eta$ mesons later decaying to the SM.
Since our lattice results show that $m_\rho > m_\eta$ for any quark mass, this process is always kinematically allowed.
\end{itemize}

On the phenomenology side, we have arrived at the following conclusions.
There is a lower limit on $m_\rho, \, m_\eta$ of a few hundred MeV from combining Higgs invisible decay constraints with bounds on the $\eta$ lifetime from nucleosynthesis.
A similar limit, $m_\rho > 280 \; \textrm{MeV}$, is required from constraints on DM self-interactions in clusters.
For larger DM masses, the parameter space is constrained by Higgs invisible decays and direct detection, implying that the scale $M$ connecting the dark sector with the Higgs field must be larger than $1-40 \; \textrm{TeV}$ depending on $m_\rho$.
We have used our lattice results to extract the $\eta$ decay constant and DM scalar form factor needed for these calculations.
At the same time, other possibilities remain for coupling our $\SU(2)$ theory to the SM (e.g. through a $Z^\prime$), which will change many of these conclusions.

\section*{Acknowledgements}

We thank Agostino Patella and Claudio Pica for their help in the initial stages of this project, in particular for sharing a version of the \verb!HiRep! software package \cite{DelDebbio:2008zf}.
The work was supported in part by the Natural Sciences and Engineering Research Council of Canada (NSERC).
Calculations were performed on the GPC machine at SciNet, as well as CEDAR and GRAHAM of Compute Canada (www.computecanada.ca). Gauge fixing, Topological Susceptibility, and Static Potential measurements were performed using GLU (https://github.com/RJHudspith/GLU).

%% appendices can go here
\appendix

%% critical mass computation
\section{Determining zero quark mass point \texorpdfstring{$m_c$}{mc} }\label{app:mc}

%%%%%%%%%%%%%%%%%%%%%%%%%%%%%%%%%%%%%%%%%%%%%%%%%%%%%%%%%%%%%%%%
\begin{figure}[h!]
\centering
\includegraphics[width=0.7\textwidth]{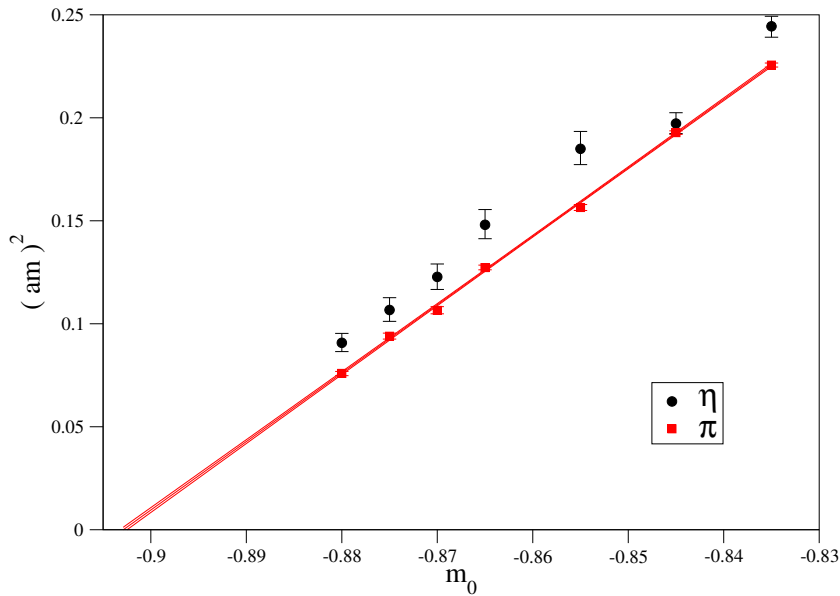}
\caption{The squared masses of the fictitious $\pi$ hadron and its comparison to the physical $\eta$ in lattice units.
}
\label{fig:PionExtrap}
\end{figure}
%%%%%%%%%%%%%%%%%%%%%%%%%%%%%%%%%%%%%%%%%%%%%%%%%%%%%%%%%%%%%%%%

Relative to the unphysical $\pi$ hadron, the physical pseudoscalar $\eta$ acquires a noticeable contribution to its mass from the U(1)$_A$ anomaly, which is clearly visible in Fig.~\ref{fig:PionExtrap}. As the 
$\pi$ is the connected part of the $\eta$, it appears as though contributions from the anomaly enter through the hairpin diagrams. Typically, we would use the non-singlett Axial Ward Identity to define the quark mass, but our theory does not have one.
Nevertheless, we use the $\pi$ to define the point of vanishing quark mass, defined as $m_c$ 
For example, it is important to know whether the $\eta$ becomes massless or if it remains massive, as expected from the anomaly.

Extrapolation of our meson masses to $m_c$ depends on the form of the extrapolation function. However, we find that a simple linear fit in $m_q$ to $m_\pi^2$ gives reasonable $\chidof=1.8$. We select where $m_\pi^2\rightarrow 0$ to be the point of our vanishing quark mass. This is consistent with another method to define this point through the topological susceptibility, described below.

Fig.~\ref{fig:PionExtrap} illustrates that the physical $\eta$ is approximately a constant shift above the $\pi$ in mass. Although this constant shift appears to be fairly small, $\frac{m_\eta^2-m_\pi^2}{\LambdaMS^2}\approx 0.25$, it does indicate that the $\eta$ remains massive in the ``chiral'' limit.

%% Topological susceptibility
\section{Topological susceptibility}
\label{app:topo}

In lattice QFT, topological charge $Q$ can be defined from the gauge fields
\begin{equation}
\begin{aligned}
Q &= \sum_x\,q(x) \,, \qquad q(x) &= -\frac{1}{32\pi^2}\epsilon_{\mu\nu\rho\sigma}{\rm Tr}\bigg[F_{\mu\nu}(x)F_{\rho\sigma}(x)\bigg],
\end{aligned}
\end{equation}
and the topological susceptibility $\chi$ can then be obtained from
\begin{equation}
\chi = \frac{\langle Q^2\rangle}{L^3T} \,.
\end{equation}
Our calculation of $F_{\mu\nu}(x)$ is the average of all four plaquettes in the $\mu-\nu$ plane that touch the point $x$, the standard clover definition.
An important issue to note with this discretisation is that the lattice values for $Q$ do not tend to be integers, due to short-distance effects which must be reduced by some smoothing procedure.
For this smoothing we will use HYP smearing \cite{Hasenfratz:2001hp}, monitoring the stability of $Q^2$ as the number of smearing iterations is increased.

We can expect for $N_f$ light quark flavors that~\cite{DiVecchia:1980yfw} 
\begin{equation}
\chi = \frac{\Sigma}{\sum_f^{N_f} \frac{1}{m_f}}\quad \underset{N_f=1}{\rightarrow} \quad \chi = \Sigma m,
\end{equation}
where $\Sigma$ is the chiral condensate. This implies that the limit $\chi \rightarrow 0$ occurs when the quark mass vanishes.

We will measure the topological susceptibility by the ``slab method''~\cite{Bietenholz:2015rsa,Aoki:2017paw}, computed on sub-volumes $V^\prime = L^3 \Delta$,
\begin{eqnarray}\label{eq:slab_method}
Q^2(\Delta) &=& \sum_{y\in V^\prime} \sum_{x\in V^\prime} \bigg\langle q(x+y,\Delta)q(y,\Delta) \bigg\rangle\; \\
&\approx& C + V\chi \left( \frac{\Delta}{T}\right).
\end{eqnarray}
For $0 < \Delta < T$, the translationally-invariant sum is best performed using convolutions over the slab.

\begin{figure}[h!]
\centering
{
\includegraphics[scale=0.28]{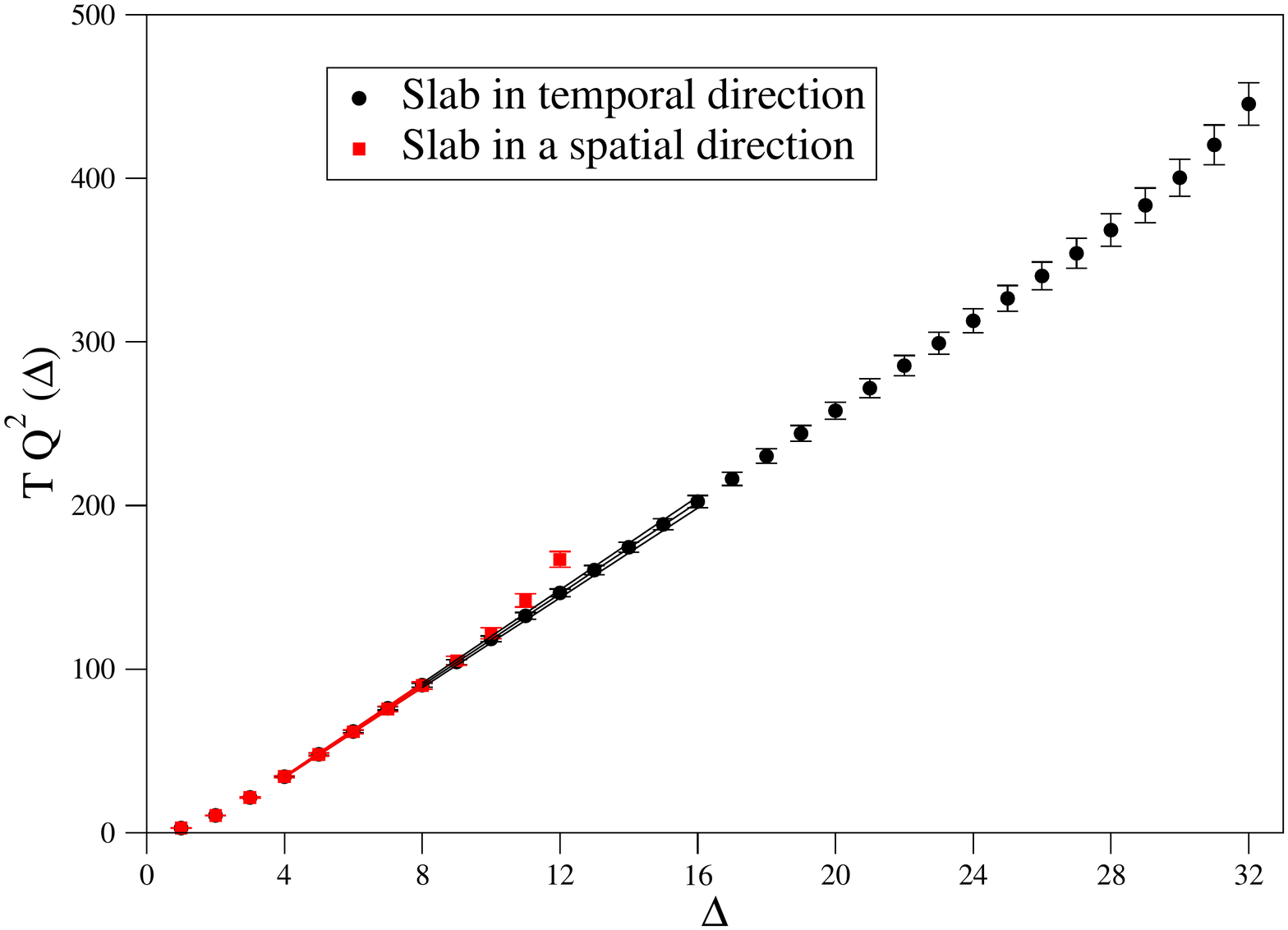}
}
\hspace{4pt}
{
\includegraphics[scale=0.28]{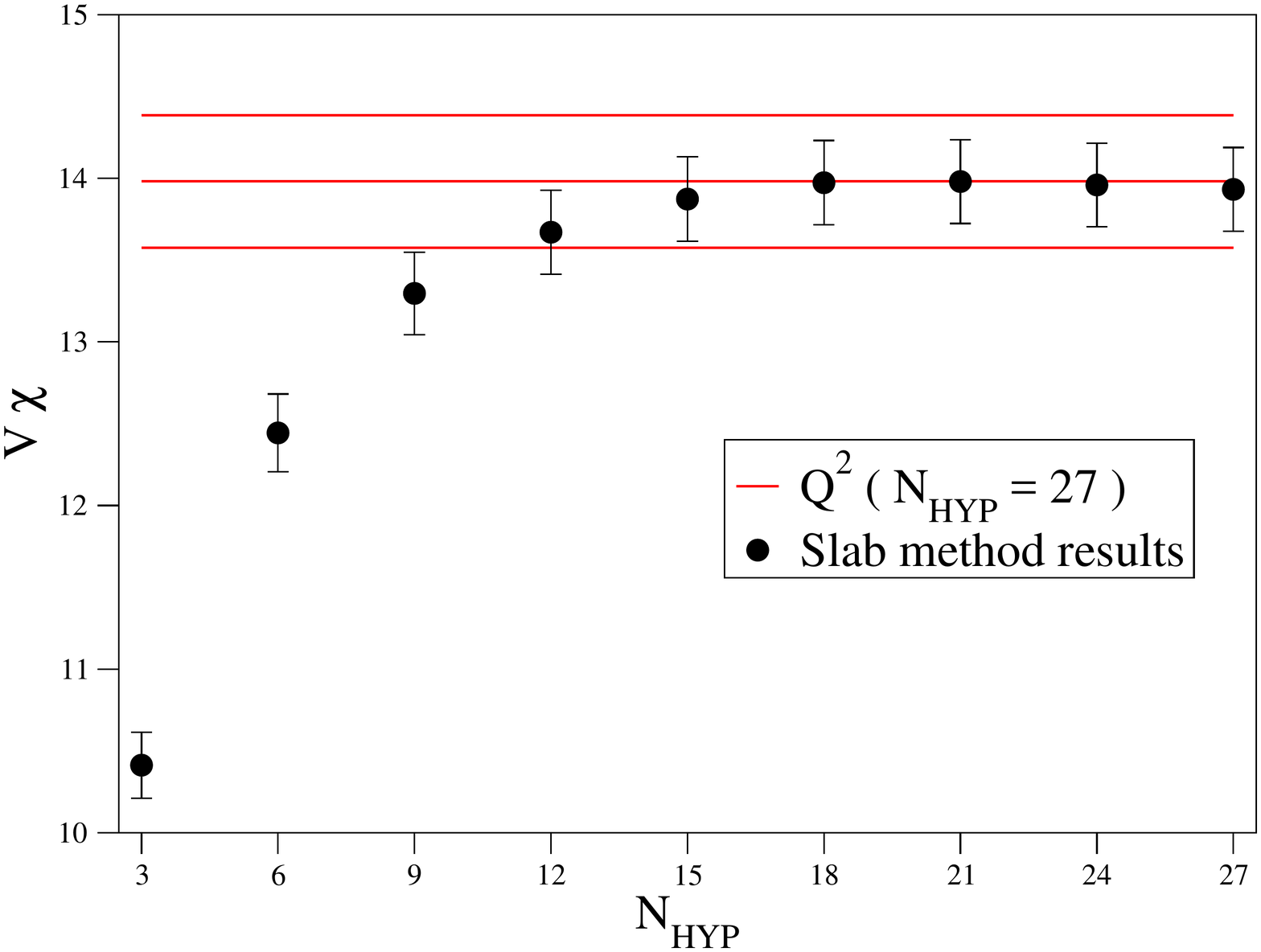}
}
\caption{Data shown here are for $m_0=-0.845$. The left panel shows the determination of $\chi$ from a linear fit to slabs along any lattice axis. The right panel compares the slab method determination (black points) to the standard measurement (1 sigma red error band).}\label{fig:slabs}
\end{figure}

\begin{figure}[h!]
\centering
\includegraphics[width=0.49\textwidth]{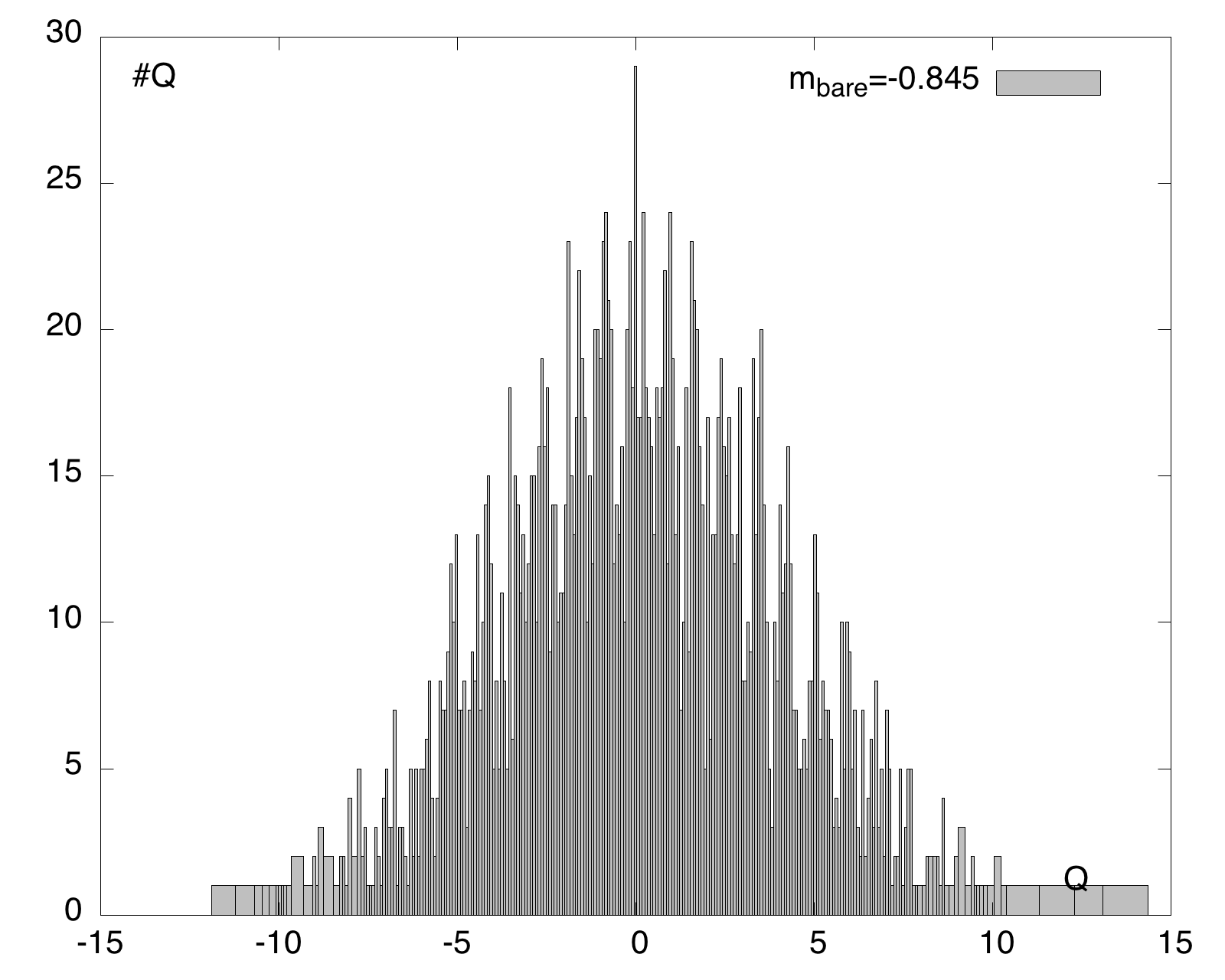}
\vspace{1ex}
\includegraphics[width=0.49\textwidth]{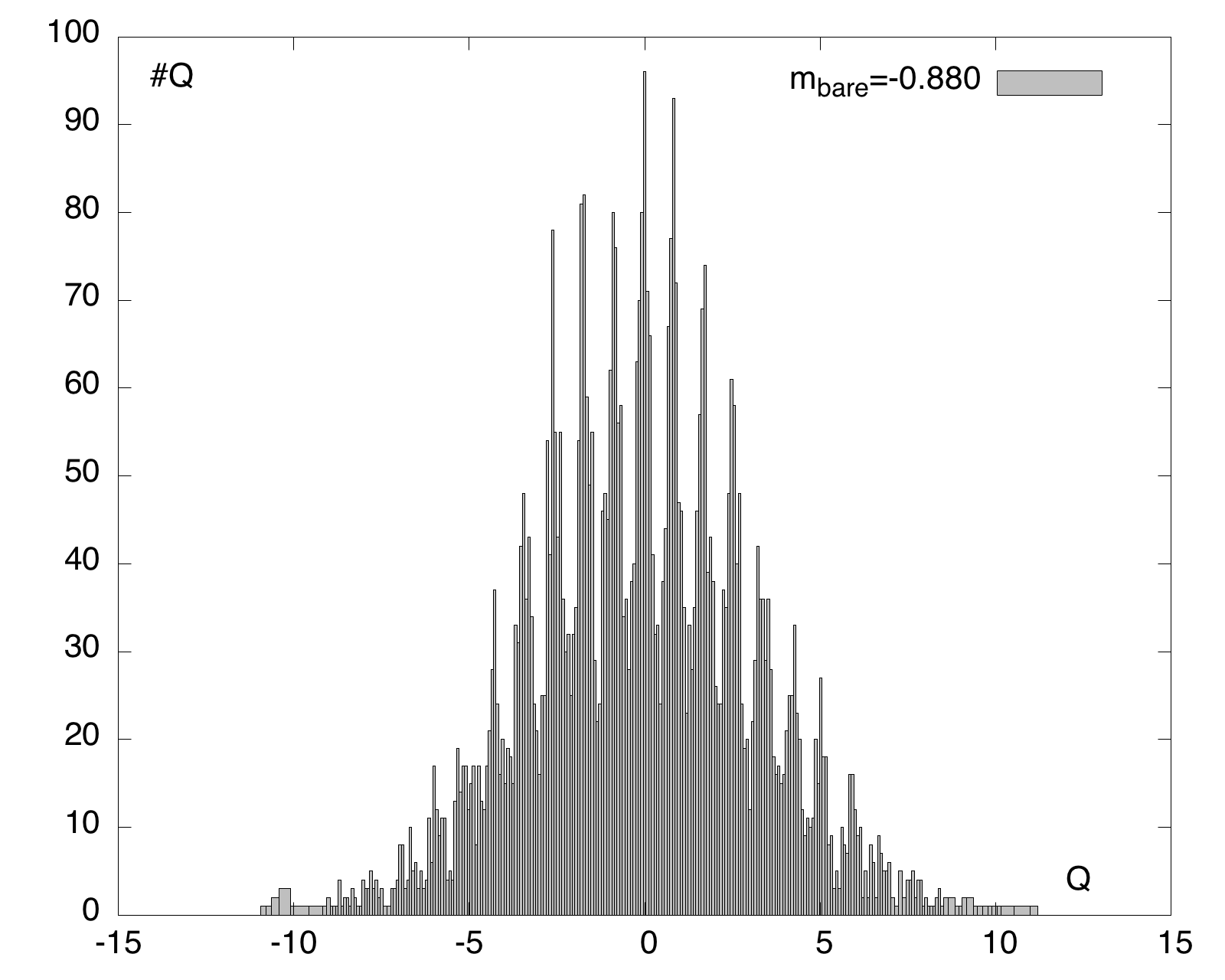}
\caption{The left panels contain histograms of topological charge of all configurations for $m_0=-0.845$ and $m_0=-0.880$ at gradient flow time $t\approx t_0$.}
\label{fig:qhist}
\end{figure}

The left panel of Fig.~\ref{fig:slabs} shows our numerical determination of the topological susceptibility.
The right panel illustrates the improvement of the slab method relative to the standard method, which is simply the slab method with $\Delta=T$.
If we fit the slab method determinations only up to $L/2$ we find good ($\simeq1.5\times$ reduction in error) statistical improvement over using the full volume determination. This indicates that the full-volume sum is noisy, and a truncated sum over a sub-volume contains less noise but still captures the relevant physics. We observe stable results after approximately 21 HYP smearing iterations.
Fig.~\ref{fig:qhist} confirms that our simulations are not getting stuck in a particular topological sector. We find that the integrated autocorrelation time for the topological charge is less than our chosen spacing for measurements in Monte-Carlo time. 

%%%%%%%%%%%%%%%%%%%%%%%%%%%%%%%%%%%%%%%%%%%%%%%%%%%%%%%%%%%%%%%%
%%%%%%%%%%%%%%%%%%%%%%%%%%%%%%%%%%%%%%%%%%%%%%%%%%%%%%%%%%%%%%%%

\begin{figure}[h!]
\centering
{
\includegraphics[scale=0.27]{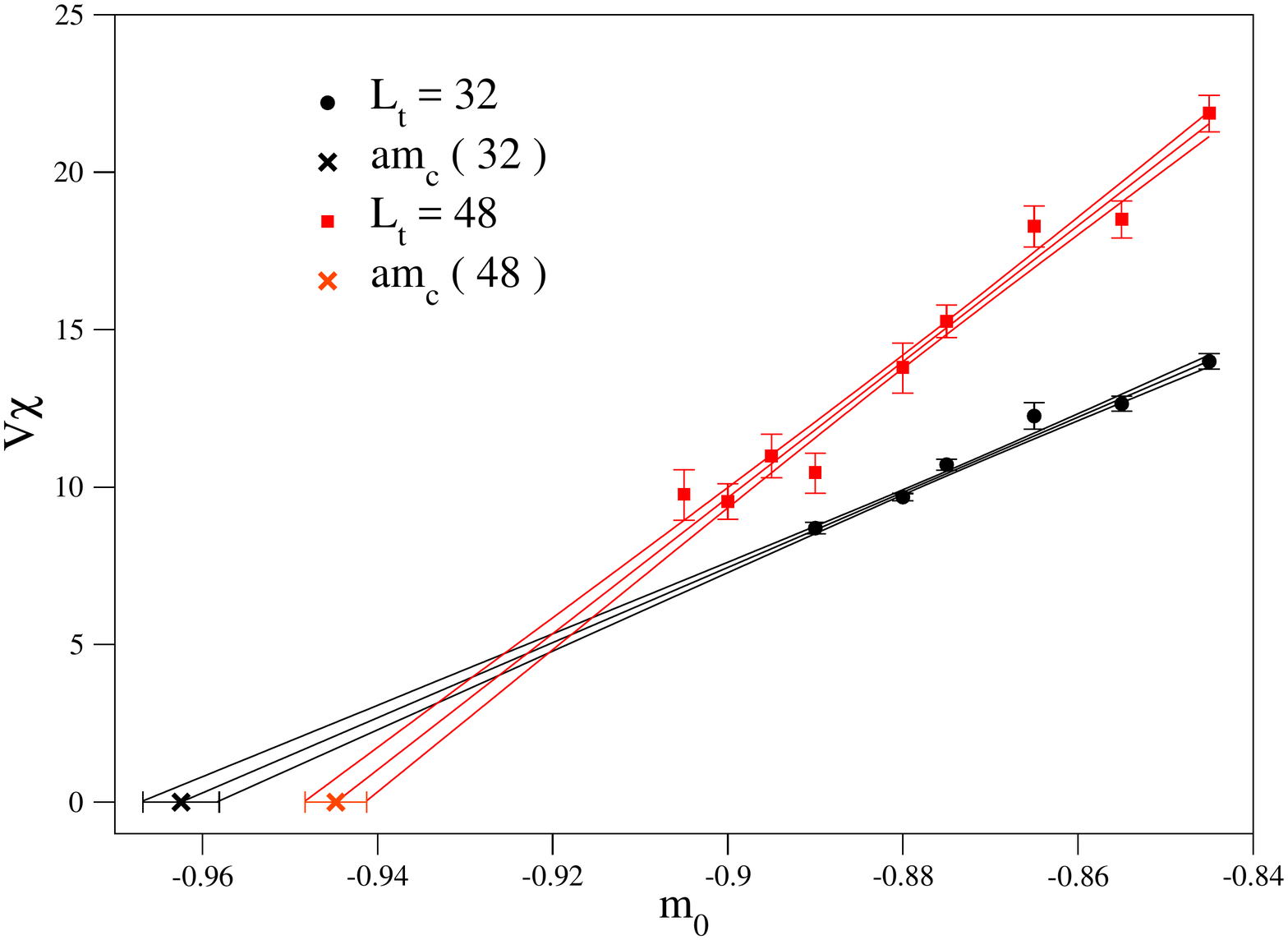}
}
\hspace{4pt}
{
\includegraphics[scale=0.27]{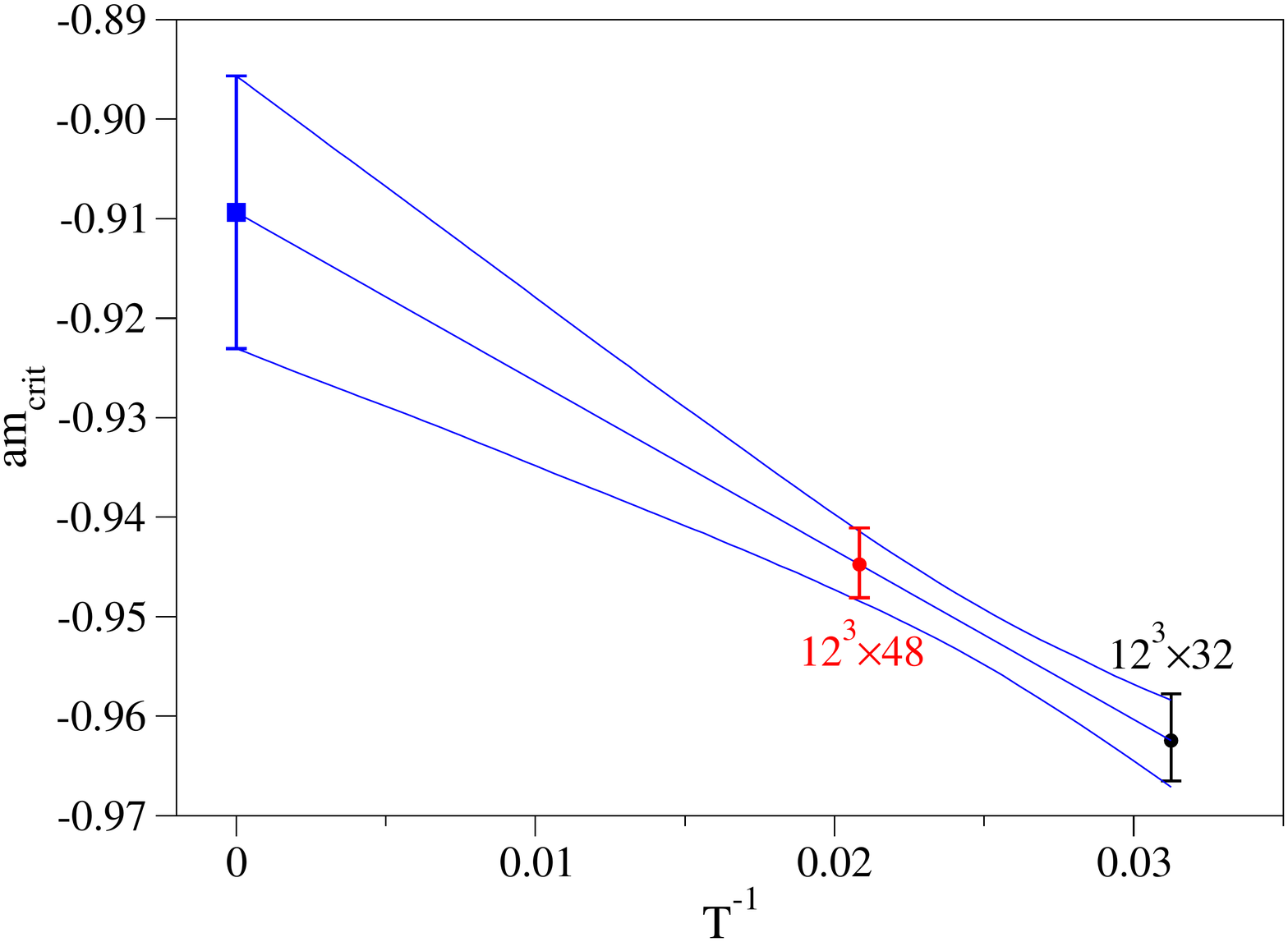}
}
\caption{The left panel shows extrapolations to vanishing topological susceptibility on $12^3\times32$ and $12^3\times48$ lattices. The right plot illustrates the infinite volume limit of this quantity.}\label{fig:slab_mc}
\end{figure}

The calculation of the topological susceptibility for a range of bare quark masses permits an extrapolation to zero as shown in the left panel of Fig.~\ref{fig:slab_mc}, representing the limit of a massless quark for that lattice volume. We have a few lighter quark masses here compared to those listed in Tab.~\ref{tab:lat_par}. These however were very difficult to invert for the meson spectrum and we suspect that they contribute large finite volume systematics to hadronic  measurements.  However, for this noisy gauge-field quantity they seem to be acceptable to use.
Repeating this procedure on a second lattice size allows an extrapolation to the limit $T\rightarrow\infty$ and our result is plotted in the right panel of Fig.~\ref{fig:slab_mc}.

From this analysis, we determine the mass at which the susceptibility vanishes as
\begin{equation}
m_c = -0.909(14) \, ,
\end{equation}
which is in good agreement with Eq.~(\ref{eq:mc}).
This consistency from two different methods is reassuring.
We will use Eq.~(\ref{eq:mc}) to define the massless limit since it has a slightly smaller error bar.

%% Setting the scale with t0 and w0
\section{The lattice scales \texorpdfstring{$t_0$ and $w_0$}{t0 and w0}
}
\label{sec:t0w0}

Our results have primarily used $\sqrt{\sigma}$ to set the physical scale of this dark matter theory, due to its direct phenomenological interpretation.
In lattice QCD calculations, however, it has become common to invoke standardized parameters named $t_0$ and $w_0$ because they can be determined much more precisely than the string tension.  We report our calculations of these quantities here to facilitate comparison with future lattice studies of this theory.

To begin, we generalize the gauge link $U_\mu(x) \to U_\mu(x,t)$ where $t$ represents the flow time.
The original, un-flowed link value is obtained at $t=0$.
The flow time does not have units of physical time, and the dimensionless quantity that emerges from a lattice simulation is $a^2t$.

Gradient flow is defined by
\begin{equation}
\frac{dU}{dt} = Z(U)U \, .
\end{equation}
$U$ is shorthand for the gauge field at a particular flow time and $Z(U)$ is chosen to be the ``force term" which is essentially the factor within the lattice action that multiplies this particular link.
The equation is solved by performing an iterated flow with (small) step size $\epsilon$,
\begin{equation}
U_{t+\epsilon} = e^{\epsilon Z(U_t)U_t^\dagger}U_t.
\end{equation}
We can use this technique to very accurately define a scale through \cite{Narayanan:2006rf,Luscher:2009eq}
\begin{equation}
G(t) = t^2 \langle F_{\mu\nu}F_{\mu\nu} \rangle , \quad G(t_0) = N/10.
\end{equation}
or through \cite{Borsanyi:2012zs}
\begin{equation}
W(t) = t\frac{d}{dt} G(t), \quad W(w_0^2) = N/10.
\end{equation}
The lattice spacing derived from these two definitions should be consistent up to discretisation effects. The factor of $N$ originates from the correct identification of the t'Hooft limit in comparison to the commonly used value of 0.3 for $\text{SU}(3)$ \cite{Ce:2016awn,DeGrand:2017gbi}.

\begin{figure}[h!]
\centering
{
\includegraphics[scale=0.27]{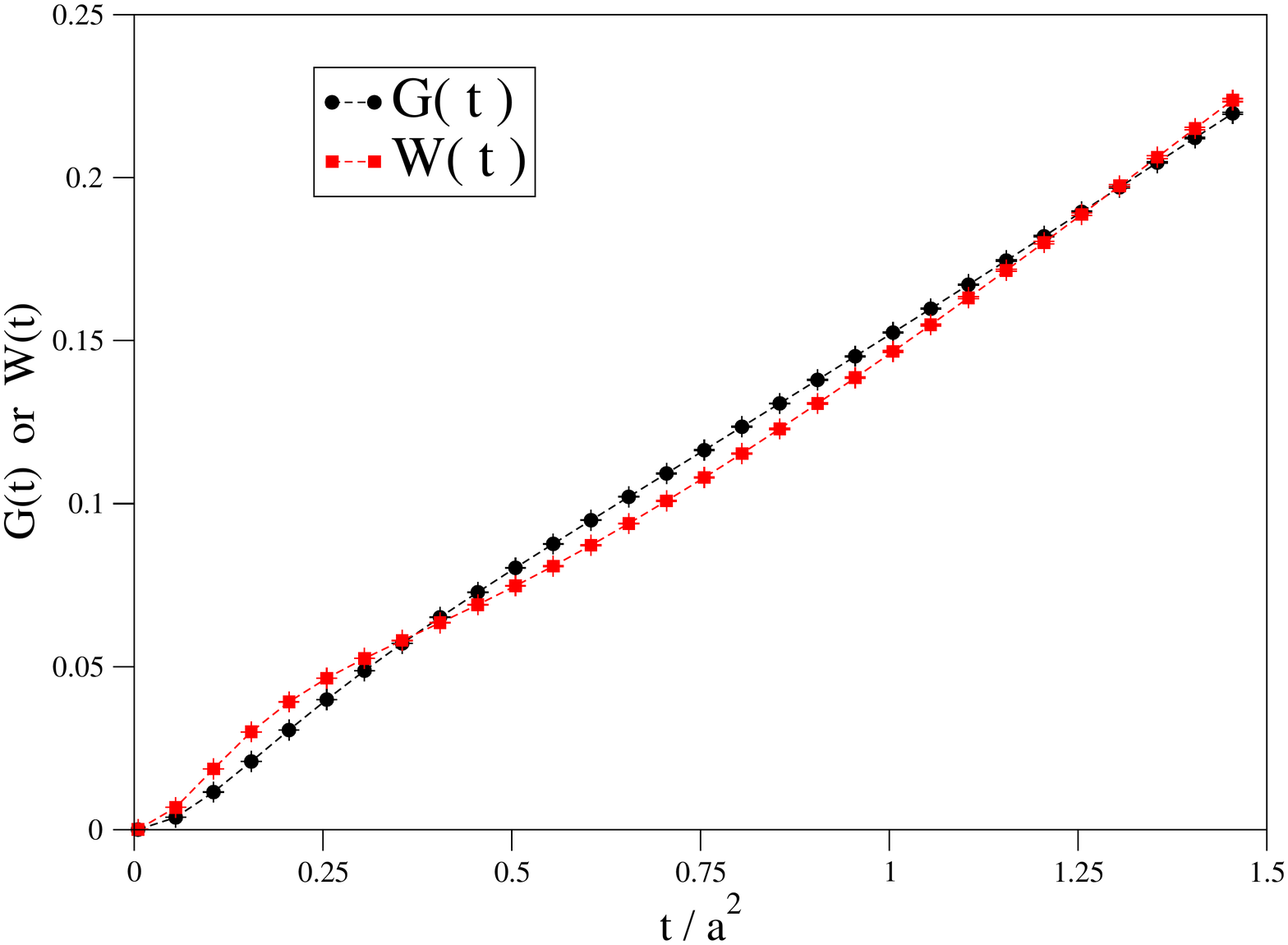}
}
\hspace{4pt}
{
\includegraphics[scale=0.27]{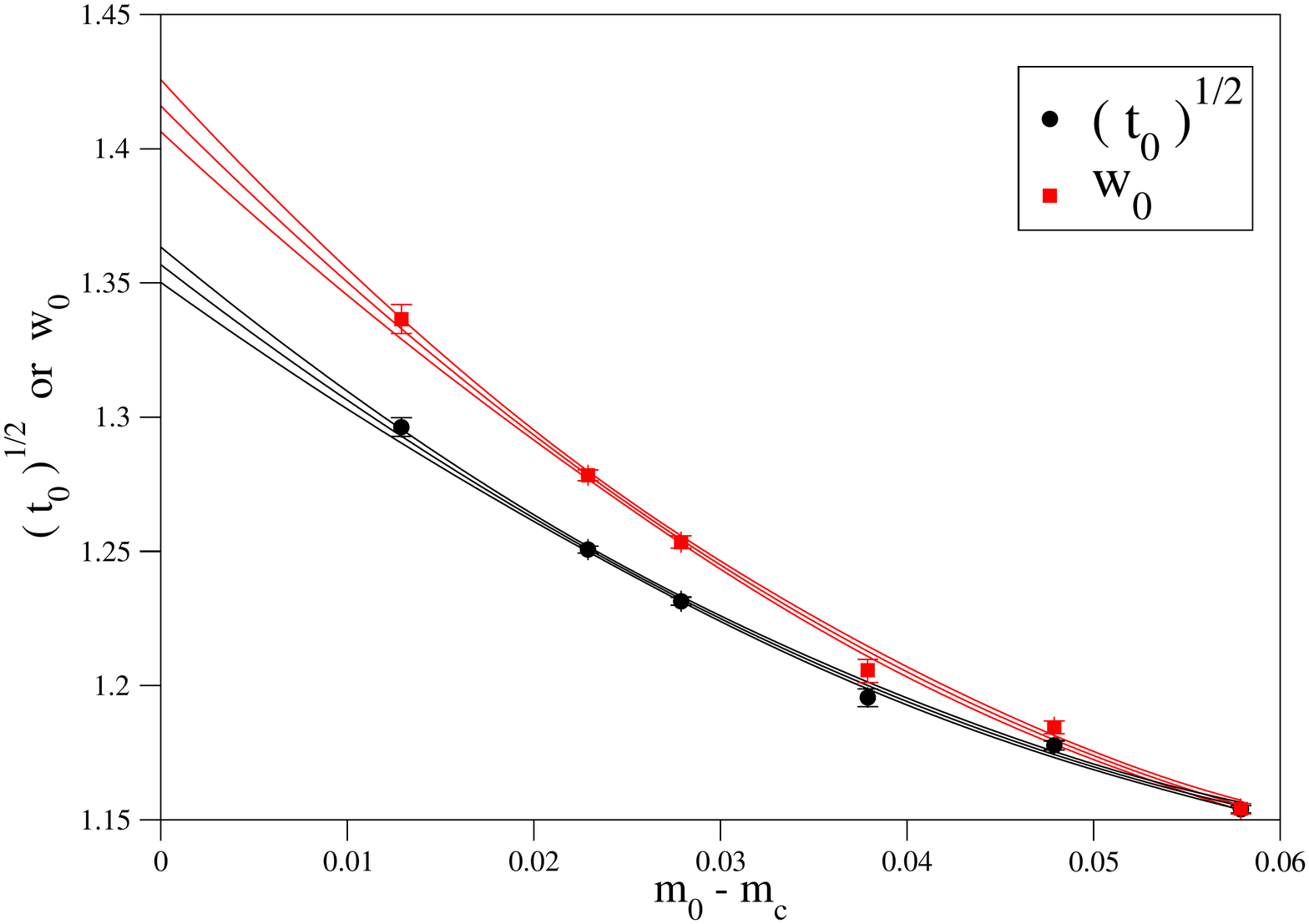}
}
\caption{Left panel: Gradient flow scale setting for the $m_0=-0.845$ ensemble.
Right panel: Results for $\sqrt{t_0}$ and $w_0$ obtained by fitting to a quadratic polynomial in $m_0-m_c$, with $\chi^2/\text{d.o.f}=2.7$ and $2.5$ respectively.}\label{fig:scales}
\end{figure}

The left panel of Fig.~\ref{fig:scales} shows our numerical results for one ensemble, and the right panel shows the fit to quark mass and the massless limit, leading to
\begin{equation}
\frac{\sqrt{t_0}}{a} = 1.357(7),\quad \frac{w_0}{a} = 1.416(10).
\end{equation}

%% Table of ensembles can go here
\section{Table of Ensembles}

\begin{table}[h!]
\caption{The bare mass $m_0$ and the number of configurations generated $N_{\rm conf}$ for the ensembles used in this work.}
\label{tab:lat_par}
\centering
\begin{tabular}{ccc}
\toprule
\multirow{2}{*}{~~~~~~$m_0$~~~~~~} & \multicolumn{2}{c}{$N_{\rm conf}$} \\
\cline{2-3}
& ~~~$T=32$~~~ & ~~~$T=48$~~~ \\
\hline
-0.105 & 175 \\
-0.305 & 244 \\
-0.405 & 200 \\
-0.505 & 170 \\
-0.605 & 645 \\
-0.705 & 660 \\
-0.755 & 374 \\
-0.805 & 248 \\
-0.815 & 233 \\
-0.835 & 1831 \\
\hline
-0.845 & 2115 & 999 \\
-0.855 & 1101 & 731 \\
-0.865 & 1354 & 602 \\
-0.875 & 2641 & 636 \\
-0.880 & 5157 & 437 \\
\botrule
\end{tabular}
\end{table}

\clearpage
\bibliography{dark}{}
\bibliographystyle{helsevier}

\end{document}